\documentstyle[12pt]{article}

\normalbaselines
\font\tbf = cmbx12

\pagebreak

\begin{document}
\indent
\vskip 2cm
\centerline{\tbf THE  CUBIC  CHESSBOARD}
\vskip 1.3cm
\centerline{Richard Kerner}
\vskip 0.8cm
\centerline{\it Laboratoire de Gravitation et Cosmologie Relativistes}
\vskip 0.2cm
\centerline{\it Universit\'e Pierre et Marie Curie - CNRS, URA 769}
\vskip 0.2cm
\centerline{\it Tour 22, 4-\`eme \'etage, Boite 142}
\vskip 0.2cm
\centerline{\it 4, Place Jussieu,} 
\vskip 0.2cm
\centerline{\it 75005 Paris, France}
\vskip 1.2cm
\centerline{\tbf Abstract}
\vskip 0.3cm
\indent
We present a survey of recent results, scattered in a series of papers that 
appeared during past five years, whose common denominator is the use of 
{\it cubic relations} in various algebraic structures.
\newline
\indent
Cubic (or {\it ternary} relations can represent different symmetries with
respect to the permutation group $S_3$, or its cyclic subgroup $Z_3$. Also
ordinary or {\it ternary} algebras can be divided in different classes with
respect to their symmetry properties. We pay special attention to the 
non-associative ternary algebra of $3$-forms (or {\it``cubic matrices''}), 
and $Z_3$-graded matrix algebras. 
\newline
\indent
We also discuss the $Z_3$-graded generalization of Grassmann algebras and
their realization in generalized exterior differential forms $d\xi$ and $d^2
\xi$, with $d^3 \xi =0$. A new type of gauge theory based on this differential 
calculus is presented. 
\newline
\indent
Finally, a ternary generalization of Clifford algebras is introduced, and
an analog of Dirac's equation is discussed, which can be diagonalized only
after taking the cube of the $Z_3$-graded generalization of Dirac's operator.
A possibility of using these ideas for the description of quark fields is
suggested and discussed in the last Section.

\newpage
\indent
\hskip 5.5cm
{\it To Andr\'e Trautman, Teacher and Friend,}
\newline
\indent
\hskip 6cm
{\it on the occasion of his $(4)^3$-th birthday}
\vskip 0.5cm
\indent
{\tbf 1. Preamble: What  I  learned  from  Andr\'e  Trautman}
\vskip 0.3cm
\indent
When the Editors of this {\it Festschrift} proposed me to write a contributed 
paper, I felt not only honored, but also deeply moved by the images of the
not so distant past that immediately emerged before my memory's eyes.
\vskip 0.2cm
\indent
Nobody has described this feeling better than Goethe \cite{Goethe}:
\vskip 0.3cm
\indent
\hskip 1.5cm
{\it Ihr naht euch wieder, schwankende Gestalten,}
\newline
\indent
\hskip 1.5cm
{\it Die fr\"uh sich einst dem tr\"uben Blick gezeigt}
\newline
\indent
\hskip 1.5cm
{\it Versuch ich wohl, euch diesmal festzuhalten ?}
\newline
\indent
\hskip 1.5cm
{\it F\"uhl ich mein Herz nach jenem Wahn geneigt ?}
\vskip 0.3cm
\indent
Yes, we were all young then, in the Warsaw of the sixties, during the short
period between the invention of the gas chambers and man's first steps on the 
Moon, when things seemed to settle down, maybe not always for the best, but 
at least in a stable way. They did not settle down for a long time, of course, 
as it always happens in history. But each generation has to discover that by 
itself.
\newline
\indent
To illustrate how young we were then it suffices to say that Andr\'e at that
time was much younger than I am now, although the age difference between us
is only eleven years, and he had been freshly promoted to the grade of
Professor at the age of 37. Life in Poland was not without shortages then
and certainly less easy than in the West. Nevertheless, the basic needs were
satisfied in general, and people still remembered that things could be much 
worse indeed.
\newline
\indent
Curiously enough, the shortages of culture and science were less acute that
those of meat and butter, a somewhat strange situation resulting from strange
ideas about human kind (\cite{Rousseau}, \cite{Fourier}, \cite{Marx}). 
Polish and Russian books were quite cheap, and most of Western books could 
be found in the Library of the Institute of Physics of Warsaw University at 
69 Hoza street. Also, with good teachers, one could get educated in modern 
Mathematics and Physics, as adequately as in the best places in the world.
\newline
\indent
Good teachers we had indeed, and Andr\'e was among the very best of them: one 
could learn the brilliance and depth of Mathematics from Krzysztof Maurin, 
the elegance and beauty of Classical Mechanics and the power of Quantum Field
Theory from Iwo Bialynicki-Birula, the universality of Geometry from
Wlodzimierz Tulczyjew, and Relativity, Gravitation and Cosmology from Andr\'e 
Trautman. I could cite yet many other excellent teachers to whom I owe my 
first and decisive training in Science.
\newline
\indent
Although when a few years later I had moved to Paris I had the privilege to 
continue my education with exceptional teachers like Andr\'e Lichnerowicz and 
Yvonne Choquet-Bruhat, this did not alter the fact that at the very foundation, 
my knowledge is based on what I have been taught during my undergraduate 
and early graduate years.
\newline
\indent
{\it "La culture c'est ce qui reste quand on a tout oubli\'e"}, said a famous
French intellectual - one of the proofs of this statement may be the fact that 
I am unable to tell now for sure whether it comes from Paul Eluard, Edouard 
Herriot or Andr\'e Malraux, - but I still remember the idea !
\newline
\indent
During the few years between 1964 and 1968, Andr\'e Trautman taught me the 
modern and unifying approach to General Relativity and Gauge Theories, 
expressed by the new means of Differential Geometry: the theory of Fibre 
Bundles and Connections, cast in an elegant and concise manner with the help 
of Cartan's exterior calculus \cite{Lich1}, \cite{Trautman1}, 
\cite{Trautman2}.
\newline
\indent
Fibre bundles were constructed as differential manifolds containing both
the {\it external} and the {\it internal} spaces. The so-called {\it external}
space was the observed four-dimensional space-time, while the internal space
was supposed to carry the {\it internal symmetries} responsible for the
{\it internal conservation laws} such as baryonic charge conservation, the 
isospin conservation, etc.  Andr\'e suggested to me that the five-dimensional 
theory of Kaluza and Klein in which the fifth dimension is identified with the 
abelian gauge group $U(1)$ should be generalized to the case of a non-abelian 
compact Lie group. We would then obtain the lagrangian describing Yang-Mills 
fields interacting with the gravitational field; the geodesic equation in the 
fibre bundle would describe the motion of the generalized ``colour'' charge 
in an external gauge field (\cite{Ker1}, \cite{Ker2},\cite{Cho},\cite{Witten}).
\newline
\indent
Later on, and in the same spirit, I constructed {\it multiple fibre bundles}, 
adding another structural group over a fibre bundle, and so on (\cite{Ker3},
\cite{Ker4}, \cite{Bertrand1}, \cite{Bertrand2}). These constructions 
introduced in a more or less natural way the Higgs fields as ``internal'' 
components of generalized gauge fields over a bundle.
\newline
\indent
In his lectures Andr\'e always underlined the facts and features that made 
clear and obvious the continuity of great ideas in science; one could see how 
the Friedmann-Lemaitre cosmological model could already be constructed in the
framework of Newtonian mechanics, with a different interpretation, of course, 
and with different observational consequences. And when one is aware of the
genuine time scale that rules the development of great ideas which stay with 
us during centuries, one is better prepared to resist the temptation of 
following short-lived fashions, even when they are declared to be the ultimate
and all-embracing ``theory of everything'' (though not accessible to 
everybody !).
\newline
\indent
Another thing that impressed me in Andr\'e's approach to science, and which
I tried to follow, was his constant quest for simplicity and elegance. The
relations between the phenomena that deserve the name of ``laws of Nature'' 
must be simple and expressed by a concise formula (\cite{Newton}), and one 
should be able to explain them as basic relations between clear concepts even 
to a person who has no wide knowledge of physics, because the deeper the 
concept is, the simpler should be the words and ideas that express it.
\newline
\indent
As an illustration to this statement I would like to recall a strange feeling 
of dissatisfaction and presence of a logical flaw, often referred to as the 
{\it contradictio in adjecto} (something nonsensical like ``a hot ice cream'' 
or ``a giant dwarf''), that often followed me while working with 
multidimensional generalizations of Einstein's theory or gauge-field theories. 
More than once it occurred to me that it is strange to use the name ``internal 
space'' (intended to be an arena of action of the group of internal symmetries 
giving rise to conservation laws unrelated to the space-time symmetries) to 
the manifold that is attached as an entity {\it exterior} to the observable 
4-dimensional space-time. One would rather expect the internal degrees of 
freedom to be encoded {\it inside} the external ones; but then the "internal 
coordinates" should be of an utterly different nature (and probably physical 
dimension as well) from the usual ones.
\newline
\indent
That is why later on I was attracted by theories in which the internal degrees 
of freedom have discrete character, like e.g. the fermionic ``coordinates'' 
in supersymmetry, or even more radically different as in the case of 
non-commutative geometries, in which the very notion of a manifold is no 
longer relevant. Together with Michel Dubois-Violette and John Madore we have 
produced some contributions in this direction, published in a series of papers 
in 1989, 1990 and 1991; important works in this area by Alain Connes, John 
Lott, Robert Coquereaux and others are by now widely known (\cite{Dubois1}, 
\cite{Dubois2}, \cite{Dubois3}, \cite{Madore1}, \cite{Connes1}, \cite{Coque}).
\newline
\indent
Since the end of the year 1990, I have started to think about 
{\it non-associative} and {\it ternary} generalizations of supersymmetry and 
the non-commutative geometries. Here again, I was led by the simple intuition 
that the smaller the scale, the stronger the interactions, so that the 
geometry at very small scales should be governed by the strongest interactions 
we know, which are apparently the {\it strong interactions} described by 
quark models.
\newline
\indent
Who knows if the magic number $3$ that seems to govern the world of quarks
is not related to the dimension $3$ of the space we are living in ? If so,
it might be worthwhile to study carefully the mathematical structures that
particular ternary symmetries display, based on the groups $Z_3$ and $S_3$ 
(resp. cyclic and general permutations of $3$ objects) instead of $Z_2$ 
(see \cite{Ker5}).
\newline
\indent
In what follows, I shall develop some arguments in favor of this new approach,
and show a few examples of its realization and consequences.
\vskip 0.4cm
\indent
{\tbf 2. Motivation for the study of ternary algebras}
\vskip 0.3cm
\indent
One of the most visible logical threads that is constantly present in modern
theoretical physics dealing with fundamental interactions is the {\it unification}. 
idea. Well known examples of this way of thinking are provided by: 
\vskip 0.2cm
\indent
a) Unified theories of Kaluza-Klein type, including the standard model of 
electroweak interactions by Weinberg and Salam, which finds its mathematical
expression in the theory of connections and curvature on fibre bundles;
\vskip 0.1cm
\indent
b) Supersymmetric theories, superposing and unifying the fields corresponding
to particles of integer and half-integer spin obeying different quantum
statistics, their mathematical framework being provided by the theory of 
superalgebras and analysis on supermanifolds;
\vskip 0.1cm
\indent
c) Quantum scattering problems and mathematical extension of group theory
to quantum groups (Hopf algebras).
\newpage
\indent
The search for unification is intimately related to the idea of 
{\it spontaneous symmetry breaking}, consisting of the conviction that we
usually observe only a partial and reduced symmetry instead of a more complete
and full symmetry that might exist among the states of the system under
consideration.
\newline
\indent
It seems possible to extend this approach of unifying symmetries to the 
algebraic structure itself, by superposing the categories of linear spaces
and the spaces of (multi) linear mappings of tensor powers of these spaces 
into themselves. One of the best ways to illustrate our ideas and to motivate 
the interest in $3$-algebras is to analyze the theory of supersymmetry from  
a slightly new point of view.
\newline
\indent
The basic ingredient of supersymmetry is the introduction of the possibility
of mixing and to superposing linearly the states corresponding to pure fermions
or bosons, which is believed to be impossible at the energies we have at our
disposal up to now. In this sense supersymmetry restores the broken symmetry,
under which the allowed transformations should never mix the fields of integer 
spin with those of half-integer spin. It is supposed that at least on some
yet unexplored energy scale the so-called {\it superfields} may appear, 
constructed from fields belonging to all possible irreducible representations 
of the Lorentz group, with coefficients belonging to the Grassmann algebra 
generated by a set of anti-commuting spinors:
\begin{equation}
\Phi(x^{\mu}, \theta^{\alpha}, {\bar{\theta}}^{\dot \beta}) := \phi_o(x^{\mu})
+ \psi_{\alpha}(x) \theta^{\alpha} + {\bar{\theta}}^{\dot{\beta}} 
{\bar{\psi}}_{\dot \beta} + W_{\mu} \sigma^{\mu}_{\alpha {\dot{\beta}}}
\theta^{\alpha} {\bar{\theta}}^{\dot \beta} +...
\end{equation}
\indent
The ``superfield'' is thus a multiplet of fields that contains a scalar $\phi_o$,
a four-vector $W_{\mu}$, a Majorana spinor $\psi_{\alpha}$ and its conjugate
${\bar{\psi}}_{\dot \beta}$, etc.; here $\theta^{\alpha}$ and 
${\bar \theta}^{\dot \beta}$ are anti-commuting variables that span the 
Grassmann algebra, and who behave like Majorana spinors with respect to 
Lorentz transformations.
\newline
\indent
Let us note at this point that spinors can be viewed as {\underline{fundamental}}
quantities here, whereas both scalar and 4-vector fields can be found as
irreducible representations in the tensor product of spinors, following the
classical decomposition formula for the representations of the Lorentz group:
\begin{equation}
D^{(\frac{1}{2} , 0)} \otimes D^{(0,\frac{1}{2})} = D^{(1,0)} \oplus 
D^{(\frac{1}{2},\frac{1}{2})}
\end{equation}
Prior to this decomposition, any tensor belonging to the product 
$D^{(\frac{1}{2},0)} \otimes D^{(0, \frac{1}{2})}$ can be identified with a 
linear operator acting on the space of spinors. This is a particular case of 
a general correspondence existing between the linear operators on a Hilbert 
space and a tensor square of this space:
\vskip 0.2cm
\centerline{ Let     $x, y \in V, \  \ A \in L(V,V)$, and let $A(x) = y$}
\vskip 0.2cm
\noindent
where $V$ is a Hilbert space, and $A$ a linear operator defined on it.
\newline
\indent
Then in a given orthonormal basis $\{ e_i \}$ , with the usual identification
of the dual basis (i.e. the basis of linear functionals on $V$) by means of
a scalar product in $H$: $e^{*k}(e_i) = \langle e_k , e_i \rangle = 
\delta^k_i$, we may write $A \sim{\displaystyle \sum_{k,i}}A^k_i e_k e^{*i}$,
so that
\newpage
\begin{equation}
A(x) = {\displaystyle \sum_{k,i}}A^k_i e_k \otimes e^{*i} ({\displaystyle
\sum_m} x^m e_m) = {\displaystyle \sum_{k,m}} A^k_m x^m e_k = y^k e_k = y
\end{equation}
\indent
The linear operator $A$ defined by its matrix elements $A^k_m$ can be also
identified with the element of the tensor product $V \otimes V$ as
\begin{equation}
A \Rightarrow  \ \ {\displaystyle \sum_{k,m}} A_{km} e_k \otimes e_m , \ \
{\rm with} \ \  A_{km} := A^k_m 
\end{equation}
\indent
In the above picture the r\^ole of the two copies of the Hilbert space that
have been used to produce general linear operators (containing the algebra
of observables) by means of the tensor product is utterly different from the
r\^ole of the third copy serving as the {\it space of states}, as may be seen
below:
\vskip 0.2cm
\centerline{$ (L(V,V) \oplus V) \sim (V \otimes V) \oplus V$}
\indent
In some sense it is also analogous with a strange and unusual summation that 
is performed on the indices of the Riemann tensor in order to obtain the Ricci 
tensor: let us recall that although all the indices of the Riemann tensor vary 
within the same range, which is the dimension of the Riemannian manifold $V_n$  
on which the metric and the Riemann tensor are defined, their nature is 
totally different. The curvature, which is a Lie algebra valued 2-form, can 
be written in a given coordinate system as:
\begin{equation}
\Omega = \Omega_{ij} dx^i \wedge dx^j = R_{ij\,m}^k E^m_k dx^i \wedge dx^j
\end{equation}
where the {\it matrices} $E^m_k$ span the basis of the $n^2$-dimensional Lie 
algebra of the linear group $GL_n(\rm{R})$ , satisfying the commutation
relations
\vskip 0.2cm
\centerline{$ E^m_k \, E^i_l \, - \, E^i_l \, E^m_k = \delta^m_l E^i_k \, -
\, \delta^i_k E^m_l$}
\vskip 0.2cm
Here again, the matrices $E^m_k$ can be put in a one-to-one correspondence
(by lowering one of the indices with the metric tensor $g_{lm}$, 
$E^m_k \rightarrow g_{lm}\,E^m_k \,= \, E_{lk}$) with the elements of the
tensor product of 1-forms, $V^* \otimes V^*$ In the definition of the Ricci
tensor, $R_{ik} = {\displaystyle{\sum_{j}}} R_{ij\,k}^j$, we perform the 
summation over the indices belonging to different realms: one comes from
the vector space, while another is a part of the {\it multi-index} labeling 
the elements of the space of {\it linear transformations (matrices)} acting
on this space.
\newline
\indent
In some sense these examples look like a discrete version of the choice 
of a {\it local section} in a fibre bundle, whose total space would be 
$V \otimes V \otimes V$.
\newline
\indent
This suggests that one could restore the full symmetry between the three
copies of the vector space $V$ by embedding $ L(V,V) \oplus V $ into the
tensor cube $V \otimes V \otimes V$, which can be then ``reduced'' or 
``projected'' in {\it six different ways} onto $L(V,V) \oplus V$.
\newline
\indent
An arbitrary element $a \in V \otimes V \otimes V$ can be written
in the basis $\lbrace e_k \rbrace$ as
\vskip 0.2cm
\centerline{$ a = {\displaystyle \sum_{k,l,m}} a_{klm} e_k \otimes e_l \otimes
e_m $,}
\noindent
which defines a one-to-one correspondence between the elements of the space
$V \otimes V \otimes V$  and the three-tensors $a_{klm}$, which we will call
also {\it cubic matrices} from now on.
\newline
\indent
It might be that one of the reasons for the non-observability of quarks is
related to the fact that they belong to this kind of ``mixed'' space, in which
there is no clear distinction between the state and the observable, both
being included in a $3$-form (or a {\it ``cubic matrix''}). 
\newline
\indent
The symmetric group  $S_3$ (of permutations of three elements) acts in an 
obvious way on the complex cubic matrices by permuting their indices. It is
therefore natural to distinguish separate subspaces of $V \otimes V \otimes V$
that provide the irreducible representations of $S_3$; e.g., there is the
subspace of the totally symmetric cubic matrices satisfying
$a_{klm} = a_{lmk} = a_{mlk} = ...$, the subspace of {\it j-skew-symmetric}
cubic matrices satisfying $b_{klm}=j\,b_{lmk}=j^2\,b_{mlk}=\bar{b_{mlk}}$, etc.
\vskip 0.2cm
\indent
Curiously enough, the most natural internal composition law that generalizes
the multiplication of ordinary matrices (or of the elements of $V \otimes V$)
is {\it ternary} and is given by the following rule:
\begin{equation}
(a \oslash b \oslash c)_{ijk} := {\displaystyle \sum_{p,q,r}} a_{ipq} b_{pjr} c_{qrk}
\end{equation}
\indent
In contrast with ordinary matrix multiplication this composition is non-
associative, in the sense that
\vskip 0.2cm
\centerline{$(a \oslash (b \oslash c \oslash d) \oslash e) \neq 
(a \oslash b \oslash c) \oslash d \oslash e \neq a \oslash b \oslash 
(c \oslash d \oslash e)$}
\vskip 0.2cm
\indent
Note that the group $S_3$ acts also on the so defined ternary algebra, so
that both actions (the permutation of factors in the product and the 
permutation of the indices in the resulting cubic algebra) can compensate
each other thus defining {\it invariant classes} in our algebra.
\newline
\indent
Such ternary products have been introduced in (\cite{Ker5} and \cite{Ker6}, 
and studied also by R. Lawrence (\cite{Lawrence}) and L. Vainerman and the
author (\cite{Vainerman}), and are, in fact, a particular case of a more 
general {\it n-fold} multiplication defined on the $n$-tensors as follows:
\[
m(a^{(1)},a^{(2)},...,a^{(n)})_{i_1i_2...i_n}=%
\sum_{j_{kr}=1(k<r)}^la_{i_1j_{12}...j_{1n}}^{(1)}a_{j_{12}i_2j_{23}...j_{2n}}^{(2)}\times ...\times 
\]
\begin{equation}
\times a_{j_{1k}...j_{k-1k}i_kj_{kk+1}...j_{kn}}^{(k)}\times ...\times
a_{j_{1n}...j_{n-1n}i_n}^{(n)}  
\end{equation}
\indent
We believe that the {\it ternary} case is exceptional because it involves the
symmetry group $S_3$ (permutations of three objects, or indices), and this 
group is the last one that possesses an exact and faithful representation in
the field of complex numbers, the next one, $S_4$, has a representation with
a double degeneracy, while starting from $S_5$ there are no representations
in $\bf C$.
\newpage
\indent
{\tbf 3. The $3$-algebra of cubic matrices: the cubic chessboard}
\vskip 0.3cm
\indent
Let us concentrate now on a more detailed study of the ternary algebra of 
complex-valued cubic matrices. Such a study seems to be particularly important
in view of the pertinence of these matrices (which are isomorphic with the
elements of $V\otimes V \otimes V$) to a possible generalization of quantum
mechanics and field theory.
\newline
\indent
Also, from a purely mathematical point of view, a ternary composition law in
a linear space over complex numbers is particularly interesting because it 
can be decomposed in an irreducible way with respect to the permutation group 
$S_3$, which is the last of the permutation groups that has a {\it faithful} 
representation in the complex plane (\cite{Ker7}). Later on we shall give 
further arguments suggesting the exceptional r\^ole of cubic matrices and 
their ternary algebra; for the time being, it suffices to draw attention to 
the fact that the cubic matrices can be visualized in three dimensions 
(like the ``Rubik's cube'', for example), and they are probably
the last case that can be still treated and analyzed in a finite time,
although even in this case the use of the computer becomes crucial.
\newline
\indent
We start by fixing the notation and conventions. Let the indices $i,k,l,m,...$ 
run from $1$ to $N$. Let the elements of a (complex-valued) matrix $a$ in a 
given basis be $a_{ikm}$. The multiplication introduced  previously is defined 
as:
\begin{equation}
(a \oslash b \oslash c)_{ijk} := {\displaystyle \sum_{p,q,r}} a_{ipq} b_{pjr}
c_{qrk}
\end{equation}
\indent
The ternary ``multiplication table'' is like a cubic chessboard with dimensions
$(N^3)\times(N^3)\times(N^3)$ , which amounts to $512$ different entries
for the case $N = 2$ and  to $27^3 = 3^9 = 19 683 $ different entries for
the case $N = 3 $ - some chessboard, indeed ! That is why in what follows we
shall restrain ourselves to the cases $N = 2$ and $N = 3$ only.
\newline
\indent
One of the natural bases in the space of cubic matrices is the set defined as:
\vskip 0.2cm
\centerline{$ e_{ikm} := 1$ at the intersection of $i$-th, $k$-th and $m$-th 
rows, $0$ elsewhere}
\vskip 0.2cm
Let us show that in the case $N = 2$ it is possible to obtain a decomposition 
of the 8-dimensional ternary algebra (as a linear space) into the direct sum 
of its three special subalgebras. In fact there are 8 matrix units in the 
whole algebra. Three of them: $e_{111},e_{222},e_{333}$ generate a subalgebra 
$Diag$ of the diagonal matrices which is evidently $S_n $-commutative. Using 
the ternary multiplication formula for the considered partial case, one can 
compute that the subalgebra generated by the matrix units $e_{112},e_{121},
e_{122}$ has a zero multiplication and consequently it is abelian. The same is 
true for the subalgebra generated by the matrix units $e_{221},e_{212},e_{211}$. 
Thus, we have a decomposition 
\vskip 0.1cm
\centerline{$Mat(2,3;C)=Diag\oplus \{e_{112},e_{121},e_{122}\}\oplus
\{e_{221},e_{212},e_{211}\}$,}
\vskip 0.1cm
in which the first summand is $S_n$-commutative and the two others are abelian 
subalgebras. This decomposition looks like a decomposition of 
$2\times 2$-matrices on the diagonal and two triangular subalgebras. But it
is not unique; one can get at least two similar decompositions: 
\[
Mat(2,3;C)=Diag\oplus \{e_{112},e_{212},e_{211}\}\oplus
\{e_{121},e_{122},e_{221}\} \ \ {\rm and} \ \
\]
\[
Mat(2,3;C)=Diag\oplus \{e_{121},e_{221},e_{211}\}\oplus
\{e_{112},e_{122},e_{212}\}.
\]
\indent
But it is another decomposition, connected with the representation properties 
with respect to the group $Z_3$ (eventually $S_3$) that will be important in
our forthcoming study of ternary algebra of cubic matrices.
\newline
\indent
Let $J \ \ $ be the cyclic permutation operator acting on cubic matrices as:
\begin{equation}
(J \, a)_{ikl} := a_{kli} ; \  \ {\rm obviously} \  \ (J^2 \,a)_{ikl} := 
a_{lik}, \  \ {\rm and} \  \ J^3 = Id
\end{equation}
\indent
In the articles (\cite{Ker5}, \cite{Vainerman}) we have introduced an 
alternative multiplication law for cubic matrices defined below:
\begin{equation}
(a*b*c)_{ikl} := {\displaystyle \sum_{pqr}} a_{piq} b_{qkr} c_{rlp}
\end{equation}
in which any {\it cyclic} permutation of the matrices in the product is 
equivalent to the same permutation on the indices:
\begin{equation}
(a*b*c)_{ikl} = (b*c*a)_{kli} = (c*a*b*)_{lik}
\end{equation}
\indent
It is easy to see that the two multiplication laws are related as follows:
\begin{equation}
a \oslash b \oslash c = (J\,a)*b*(J^2 \,c) ,
\end{equation}
\noindent
and neither of the two is associative.
Let us denote by $j$ the cubic root of unity, $j = e^{\frac{2\pi i}{3}}$;
we have $j + j^2 + 1 = 0$, and $\bar{j} = j^2$.
\newline
\indent
The complex square $N \times N$-matrices can be divided into subspaces 
with particular representation properties with respect to the group of 
permutations $S_2$ (isomorphic with $Z_2$), thus defining {\it symmetric}, 
{\it anti-symmetric}, {\it hermitian} and {\it anti-hermitian} matrices: 
let $T$ be the transposition operator, $(Ta)_{ik} = a_{ki}$; then
we can define the aforementioned types of matrices as the ones that have the
following transformation laws under the action of $T$ :
\newline
\indent
\centerline{$Ta = a,\ \ \ \ Ta = - a,\ \ \ \ Ta = \bar{a},\ \ \ \ Ta = - \bar{a}$}
which gives in index notation the usual definitions:
\newline
\centerline{$ a_{ik} = a_{ki}, \ \ \ \ a_{ik} = - a_{ki}, \ \ \ \  
a_{ik} = {\bar{a}}_{ki}, \ \ \ \ a_{ik} = - {\bar{a}}_{ki}$,}
\vskip 0.2cm
Similarly, the complex cubic matrices can be divided into classes according 
to the representations of the group $S_3$. With $J$ defined as above (9) and
$T$ the operator of odd transposition, $(Ta)_{ikm} = a_{mki}$, we can define 
cubic matrices with the following non-equivalent representation properties 
under the action of the operators $J$ and $T$:
\begin{equation}
Ja = a, \ \ Ta = a;  \ \ \ \ Ja = j\,a,\ \ Ta = a; \ \ \ \ Ja = j^2 \,a,\ \ 
Ta = a
\end{equation}
and
\begin{equation}
Ja = a, \ \ Ta = \bar{a}; \ \ \ \ Ja = j\,a, \ \ Ta = \bar{a} ; \ \ \ \ 
Ja = j^2 \,a, \ \ Ta = \bar{a}
\end{equation}
\indent
From now on we shall concentrate on the class of matrices displaying 
well-defined properties with respect to the group of cyclic permutations $Z_3$
only, i.e. supposing that there is no particular relation between $a$ and $Ta$.
\newline
\indent
This type of decomposition is important in the analysis of the possible
representations of ternary algebras of cubic matrices in terms of associative 
matrix algebras. We shall follow the well known example of Ado's theorem for 
finite-dimensional Lie groups, which states that for such groups an associative 
{\it enveloping } algebra can be found, such that the skew-symmetric, 
non-associative composition law satisfying the Jacobi identity can be 
faithfully represented by a {\it commutator} of the corresponding elements.
\newline
\indent
Although at this stage we are don't know if an analogue of the Jacobi identity 
exists for ternary algebra of cubic matrices, we shall show that at least for 
the simplest cases, certain ternary algebras with a non-associative
composition law displaying particular symmetries can be represented in the
algebra of associative matrices. Let us decompose the algebra of cubic 
matrices into the direct sum of the following linear subspaces:
\vskip 0.2cm
\noindent
{\it Diagonal,} containing N diagonal cubic matrices $\omega^{(k)}$ : 
\vskip 0.2cm
\centerline{$\omega^{(k)}_{kkk} = 1$, all other elements $= 0$;}
\vskip 0.2cm
\noindent
{\it Symmetric,} containing $(N^3-N)/3$ {\it traceless, totally symmetric} 
cubic matrices  
\vskip 0.2cm
\centerline{$\pi^{(\alpha)}_{klm} = \pi^{(\alpha)}_{lmk} = \pi^{(\alpha)}_{mkl}
\, , \ \  \ \  \alpha = 1, 2, .., (N^3-N)/3$}
\vskip 0.2cm
\noindent
{\it j-Skew-symmetric,} containing $(N^3-N)/3$ cubic matrices satisfying
\vskip 0.2cm
\centerline{$\rho^{(\alpha)}_{klm} = j\,\rho^{(\alpha)}_{lmk} = j^2\,
\rho^{(\alpha)}_{mkl}$;}
\vskip 0.2cm
\noindent
and {\it $j^2$-Skew-symmetric,} containing $(N^3-N)/3$ cubic matrices 
satisfying
\vskip 0.2cm
\centerline{$\kappa^{(\alpha)}_{klm} = j^2\, \kappa^{(\alpha)}_{lmk} = 
j\, \kappa^{(\alpha)}_{mkl}$.}
\vskip 0.2cm
\indent
Only the diagonal matrices form a $3$-subalgebra with respect to the ternary
multiplication law; 
\begin{equation}
\omega^{(k)}\,* \omega^{(l)}\, * \omega^{(m)} = 0 \ \ {\rm if} \ \ \ \ 
k \neq l \neq m \ \ \ \ {\rm and} \ \ 
\omega^{(k)}\,* \omega^{(k)}\,* \omega^{(k)} = \omega^{(k)}
\end{equation}
\indent
This $3$-subalgebra is associative and commutative and is easily represented
by ordinary (square) matrices $\omega^{(k)}$, whose only non-vanishing element 
$1$ is found at the intersection of the k-th line with the k-th column.
\newline
\indent
With eight independent generators the ternary algebra's multiplication table
is also a {\it cubic} array, and in order to define it completely we must
display as many as $8 \times 8 \times 8 = 512$ different ternary products.
Because of the non-associativity of ternary law, it is impossible to find
a realization of these multiplication rules by means of a set of finite  
$n \times n$ matrices. 
\newline
\indent
This situation is not new indeed, and could be observed in the case of binary
non-associative algebras. The well known Ado's theorem states that a class
of finite dimensional non-associative algebras with particular symmetry of 
the composition law, $\{X,Y\} = - \{Y,X\}$ and satisfying the Jacobi identity 
(Lie algebras) can always be represented by a subset of some bigger 
{\it associative} algebra, called the enveloping algebra.
\newline
\indent
Let us show on a simple example of $2 \times 2 \times 2$ cubic matrices that 
a representation in the associative binary algebra of $2 \times 2$ ordinary 
matrices can be found provided that the ternary composition law is endowed 
with a particular symmetry that generalizes the skew symmetry of the ordinary 
Lie algebra.
\newline
\indent
In the multiplication table for the cubic matrices in the particular basis
of $\omega^{(k)}, \pi^{(\alpha)}, \rho^{(\beta)}$ and $\kappa^{(\gamma)}$
it is difficult to find any subalgebras except for the obvious ``central'' one
containing the $\omega^{(k)}$. Usually a $3$-product of three matrices will
decompose into a linear combination of the matrices belonging to various
symmetry types, e.g.
\vskip 0.2cm
\centerline{$ \rho^{(1)} \,* \rho^{(1)} \,* \rho^{(2)} = \omega^{(1)} -
\frac{1}{3} \pi^{(2)} + \frac{2}{3} \, j^2 \, \rho^{(2)} - \frac{1}{3}\, j\,
\kappa^{(2)}$ , etc.}
\vskip 0.2cm
The situation changes if we introduce a new composition law that follows the 
particular symmetry of the given type of cubic matrices. For example, let us 
define:
\begin{equation}
\{\rho^{(\alpha)} , \rho^{(\beta)} , \rho^{(\gamma)} \} := \rho^{(\alpha)} *
\rho^{(\beta)} * \rho^{(\gamma)} + j\, \rho^{(\beta)} * \rho^{(\gamma)} *
\rho^{(\alpha)} + j^2 \, \rho^{(\gamma)} * \rho^{(\alpha)} * \rho^{(\beta)}
\end{equation}
Because of the symmetry of the ternary {\it j-bracket} one has
\vskip 0.2cm
\centerline{$ \{ \rho^{(\alpha)}, \rho^{(\beta)} , \rho^{(\gamma)} \}_{ikm}
= j \{ \rho^{(\alpha)} , \rho^{(\beta)} , \rho^{(\gamma)} \}_{kmi} $,}
\vskip 0.2cm
\noindent
so that it becomes obvious that with respect to the {\it j-bracket} 
composition  law the matrices $\rho^{(\alpha)}$ form a ternary subalgebra.
Indeed, we have
\begin{equation}
\{ \rho^{(1)} , \rho^{(2)} , \rho^{(1)} \} = - \rho^{(2)} \, ; \ \ \ \ 
\{ \rho^{(2)} , \rho^{(1)} , \rho^{(2)} \} = - \rho^{(1)} \, ;
\end{equation}
all other combinations being proportional to the above ones with a factor $j$
or $j^2$, whereas the j-brackets of three identical matrices obviously
vanish.
\newline
\indent
Our aim is to find the simplest representation of this ternary algebra in
terms of a {\it j-commutator} defined in an associative algebra of matrices
$M_2({\tbf C})$ as follows:
\begin{equation}
\lbrack A, B, C \rbrack := A B C + \, j \, B C A + \, j^2 \, C A B
\end{equation}
\indent
It is easy to see that the trace of any j-bracket of three matrices must
vanish; therefore, the matrices that would represent the cubic matrices
$\rho^{(\alpha)}$ must be traceless. Then it is a matter of simple exercise
to show that any two of the three Pauli sigma-matrices divided by $\sqrt{2}$
provide us with a representation of the ternary j-skew algebra of the 
$\rho$-matrices; e.g.
$$ \sigma^1 \sigma^2 \sigma^1 + j\,\sigma^2 \sigma^1 \sigma^1 +
j^2 \, \sigma^1 \sigma^1 \sigma^2 = - 2 \, \sigma^2, \ \ 
\sigma^2 \sigma^1 \sigma^2 + j\, \sigma^1 \sigma^2 \sigma^2 
+ j^2 \, \sigma^2 \sigma^2 \sigma^1 = - 2 \, \sigma^1 $$
Thus, it is possible to find a representation in the associative algebra of
finite matrices for the non-associative j-bracket ternary algebra. A similar 
representation can be found for the two cubic matrices $\kappa^{(\alpha)}$ 
with the $j^2$-skew bracket.
\newline
\indent
If there  exists an analogue of the Jacobi identity, it cannot contain the
{\it double j-brackets}. As a matter of fact, we have been able to prove
that there are no non-trivial solutions to the equation containing the forty
non-redundant double j-brackets like $\lbrack A, \lbrack B, C, D \rbrack , E
\rbrack $. Therefore, in order to produce a non-trivial analogue of the
Jacobi identity for ternary algebras, we should find an identity involving
{\it seven} different entities, like 
\vskip 0.2cm
\centerline{$\lbrack A, \lbrack B , \lbrack C, D, E \rbrack , F \rbrack G
\rbrack$, and $\lbrack \, \lbrack A, B, C \rbrack \,D \, \lbrack E, F, G 
\rbrack \rbrack$, etc.}
\vskip 0.2cm
\indent
It is also worthwhile to note that the ordinary Lie algebras with the 
skew-symmetric composition law can be found in the representation of the
ternary j-bracket algebra in the associative algebra, provided the latter one
is endowed with a central (unit) element. Indeed, we have:
\begin{equation}
\lbrack A, {\bf 1} , C \rbrack = A\, {\bf 1} \, C + \, j\,{\bf 1} \, C \, 
A + \, j^2 \, C \, A \, {\bf 1} = A\,C + (j + j^2)\, C \, A = A\,C - C\, A
\end{equation}
\indent
The fact that Pauli matrices did appear in a quite natural way is encouraging.
It suggests that although we start here from a ternary algebra with 
$j$-skew $3$-commutator, more familiar notions such as the Lorentz group and
spin can be encoded in some way in this unusual rules, and appear sooner or
later as secondary features of a purely algebraic theory.
\newline
\indent
The following exercise reinforces this hope.
\newline
\indent
A natural question to ask now concerns the nature of all the automorphisms
of this simple ternary algebra. The most general homogeneous transformation
of the cubic matrices $\rho^{(\alpha)}$ involves all their indices:
\begin{equation}
{\tilde \rho}^{(\alpha)}_{ikm} = \Lambda^{\alpha}_{\beta}\, U^p_i \, U^r_k \,
U^s_m \, \rho^{(\beta)}_{prs} ,  \ \ \ \ \alpha, \beta, i, k, ...=1,2.
\end{equation}
with (invertible) matrices $\Lambda^{\alpha}_{\beta}$, $U^p_i$ chosen in such
a way that the ternary relations between the transformed cubic matrices
${\tilde \rho}^{(\alpha)}$ remain the same as defined above.
\newline
\indent
Let us show that even in a simplified case when we choose $U^p_q = \delta^p_q$,
the condition of invariance of the ternary algebra leads to non trivial
solutions for the group of matrices $\Lambda^{\alpha}_{\beta}$. As a matter
of fact, we get the following system of equations for $\Lambda^{\alpha}_%
{\beta}$ :
\begin{equation}
\Lambda^1_1 (\Lambda^2_2 \Lambda^1_1 - \Lambda^1_2 \Lambda^2_1) = \Lambda^2_2; \ \ \ \
\ \ \ \
\Lambda^1_2 (\Lambda^2_1 \Lambda^1_2 - \Lambda^1_1 \Lambda^2_2) = \Lambda^2_1,
\ \ {\rm and}
\end{equation}
\begin{equation}
\Lambda^2_2 (\Lambda^1_1 \Lambda^2_2 - \Lambda^2_1 \Lambda^1_2) = \Lambda^1_1;
\ \ \ \  \ \ \ \
\Lambda^2_1 (\Lambda^1_2 \Lambda^2_1 - \Lambda^2_2 \Lambda^1_1) = \Lambda^1_2
\end{equation}
\indent
from which follows that $[det(\Lambda)]^2 = 1$, so that either
\begin{equation}
det(\Lambda) = 1 , \ \  \ \ {\rm and} \ \ \ \ \Lambda^1_1 = \Lambda^2_2, \ \
\Lambda^1_2 = - \Lambda^2_1 , \ \ {\rm or} 
\end{equation}
\begin{equation}
det(\Lambda) = - 1 , \ \ \ \ {\rm and} \ \ \ \ \Lambda^1_1 = - \Lambda^2_2,
\ \ \Lambda^1_2 = \Lambda^2_1.
\end{equation}
\indent
This group has two disjoint components; the simply connected component of the
unit element is a subgroup, whereas the second component can be obtained from 
the first one by multiplication by the $2 \otimes 2$ matrix $diag(1, -1).$
\newline
\indent
The simply connected subgroup is an abelian, (real) two-dimensional Lie group 
of matrices whose general form is
\begin{equation}
\pmatrix{a & b \cr -b & a} , \ \ {\rm with} \ \ a, \ \ b  \ \ 
{\rm complex \ \ numbers \ \ satisfying} \ \ a^2 + b^2 = 1
\end{equation}
which can be decomposed into a simple product of two matrices:
\begin{equation}
\pmatrix{\cosh \psi & i \sinh \psi \cr -i \sinh \psi & \cosh \psi } 
\pmatrix{\cos \phi & \sin \phi \cr - \sin \phi & \cos \phi }
\end{equation}
\indent
This group is easily identified as the simple product of Euclidean rotations
and translations. It can be realized as the isometry group of a cylindrical 
Minkowski space parametrised with two variables 
$\tau$ and $\phi$, $\lbrack 0 \leq \phi \leq 2 \pi \rbrack \times \lbrack 
- \infty \leq \tau \leq \infty \rbrack $ , with one ``boost'' and one angular
translation. When embedded in a many-dimensional Minkowski space, this object
looks like a motionless closed string. This invariance group reduces to $U(1)$ 
if we impose the reality condition on the matrices $\rho{(\alpha)}$ requiring 
that $\rho^{(\alpha)}_{ikl} ={\bar \rho}^{(\alpha)}_{lki}$.
\newline
The ternary algebra of complex cubic matrices in three dimensions,
$ Mat(3,3,{\tbf C}) $, has a very rich structure; its multiplication table 
(in three dimensions, too) can be visualized as a cubic matrix with 
$27 \times 27 \times 27 = 3^9$ entries. Here again, subsets displaying a 
particular $Z_3$-symmetry can be defined, containing eight independent 
matrices each, so that the whole algebra of $27$ independent matrices 
decomposes as $M = Diag \oplus M_0 \oplus M_1 \oplus M_2 $.
\newline
\indent
Let us denote these cubic matrices by: $ O^{(a)}_{bcd} $ (the diagonal  part);
$R^{(A)}_{abc}$, with $A = 1,2,...8$ , and $a,b = 1,2,3 $ spanning the subset 
$M_1$; $K^{(A)}_{abc}$ spanning the subset $M_2$, and $P^{(A)}_{abc}$   
spanning the totally $Z_3$-symmetric traceless subset $M_0$. The cubic
matrices denoted by capital Latin letters display the same symmetries as their
prototypes belonging to $Mat(2,3,{\tbf C})$ denoted by the corresponding
Greek letters $\omega$,$\rho$,$\kappa$ and $\pi$. It is easy to see that the 
component $M_1$ containing the matrices $R^{(A)}$ satisfying
\vskip 0.2cm
\centerline{$R^{(A)}_{abc} = j\,R^{(A)}_{bca} = j^2\, R^{(A)}_{cab}$}
\vskip 0.1cm
\noindent
consists of three two-dimensional ternary subalgebras, each of them isomorphic 
with the algebra of $\rho$-matrices shown above. The three subalgebras are
spanned by (we just give the only non-vanishing elements):
\vskip 0.2cm
\centerline{$\{ R^{(1+)}_{232} \, , R^{(1-)}_{323} \} \, ; \ \ \{ R^{(2+)}_{313} \, ,
R^{(2-)}_{131} \} \, \ \ {\rm and} \ \ \{ R^{(3+)}_{121} \,,R^{(3-)}_{212} \}$}
\vskip 0.2cm
besides, there are two more independent generators,
\vskip 0.2cm
\centerline{$ R^{(7)}_{123}$ and $R^{(8)}_{321} $}
\vskip 0.1cm
\indent
This situation is similar to the one observed in the examples of the Lie 
algebras $su(2)$ and $su(3)$, where the algebra $su(2)$ can be embedded in 
three different ways in the algebra $su(3)$. It is also clear that among the
automorphisms of the ternary algebra spanned by $R^{(A)}$, with the $j$-skew 
ternary commutator, we will find three copies of the automorphisms of the
simple ternary algebra of $\rho^{(\alpha)}$ cubic matrices, which means that
we shall have {\it three} independent Lorentzian boosts, and {\it three}
independent rotations of a plane, which is exactly what is needed to generate 
the $6$-parameter Lorentz group in 4-dimensional space-time. Similar 
observation can be made concerning the cubic matrices $K^{(A)}$.
\newline
\indent
This does not exclude the possibility of finding other interesting subgroups 
in the group of automorphisms of ternary relations between the cubic matrices 
$R^{(A)}$ or $K^{(A)}$, e.g. the group $SU(3)$ in its adjoint representation, 
althoughit may be intertwined with the elements of the Lorentz group in a very 
tricky way.
\newline
\indent
To find a maximal ternary subalgebra of $M_1 \subset Mat(3,3;{\tbf C})$ that
can be represented in a finite associative algebra with the $j-skew$ 
commutator as the composition law is not an easy task, and we don't know the
full answer to this problem. However, the fact that the traceless part of 
the $3$-algebra of cubic matrices splits naturally into three equal parts
suggests that its representation by means of an associative eneveloping
algebra can be naturally $Z_3$-graded, with three grades $0,1,2$ adding
up modulo $3$. Such algebras are also very interesting, and we were able to
investigate them to some extent; some of the results are presented in the
following Sections.
\vskip 0.4cm
\indent
{\tbf 4. $Z_3$-graded associative algebras}
\vskip 0.3cm
\indent
The simplest case of an associative $Z_3$-graded algebra is provided by the
algebra of complex $3 \times 3$ matrices, which is divided into the following
three linear subspaces: ${\cal{M}}_3 (\tbf{C})={\cal{A}}_0 \oplus {\cal{A}}_1 
\oplus {\cal{A}}_2 $, with
\begin{equation}
\pmatrix{\alpha & 0 & 0 \cr 0 & \beta & 0 \cr 0 & 0 & \gamma} \in {\cal{A}}_0; 
\ \ \ \ \pmatrix{0 & \alpha & 0 \cr 0 & 0 & \beta \cr \gamma & 0 & 0 } \in  
{\cal{A}}_1 ; \ \ \ \ \pmatrix{0 & 0 & \gamma \cr \alpha & 0 & 0 \cr 0 & 
\beta & 0} \in {\cal{A}}_2 \ \ \ \, \alpha, \beta, \gamma \in {\tbf C}.
\end{equation}
One easily checks that if the elements $a^{(k)}$ and $b^{(m)}$ belong to the
subspaces ${\cal{A}}_k$ and ${\cal{A}}_m$ respectively, ($k,m = 0,1,2$), then
their matrix product belongs to ${\cal{A}}_{(k+m)mod(3)}$. This algebra may 
be viewed upon as a model of non-commutative $Z_3$-graded geometry, 
generalizing the $Z_2$-graded matrix algebra used in the models of elementary 
interactions based on the noncommutative $Z_2$-graded geometry, (\cite{Dubois1}, 
\cite{Dubois2}, \cite{Dubois3}, \cite{Connes1}, \cite{Coque}, \cite{Chams}, 
\cite{Scheck}).
\newline
\indent
Let us introduce the following $Z_3$-graded commutator:
\begin{equation}
[ A,\,B ]_{Z_3} = A\,B - j^{ab} B\,A, \ \ {\rm with} \ \ 
a = grad(A), \ \ b = grad(B).
\end{equation}
Denoting the $Z_3$-commutator of element $A$ with any other element $B$ by
$Der_A (B)$, (both of them having the well defined $Z_3$-grade), it is easy 
to check the $Z_3$-graded Leibniz rule in $\cal{A}$:
\begin{equation}
Der_A\,(B\,C) = [Der_A\,(B)]\,C + j^{ab}\,B\,[Der_A\,(C)]
\end{equation}
However, these derivations do not form a ($Z_3$-graded) Lie algebra, because 
the iterated $Z_3$-commutator $[A,\,[B,\,C]]$ cannot be expressed as a linear 
combination of two $Z_3$-commutators  $[[A,\,B],\,C]$ and $[B,\,[A,\,C]]$ 
(i.e. the $Z_3$-graded analogue of the Jacobi identity does not exist here).
\newline
\indent
The derivations are naturally divided into three distinct classes, following
their $Z_3$-grade. The derivations of $Z_3$-grade $1$ and $2$ are {\it cubic
nilpotent}: if $grad(A) = 1 \ \ {\rm or} \ \ 2$,
\begin{equation}
(Der_A)^3\,B = 0 \ \ { \rm for \ \ any \ \ } \, B
\end{equation}
for example, the simple calculus for k=1 shows quite immediately that
$${D_1}^3 B := {[ A ,{[ A , {[ A , B ]}_{Z_3}]}_{Z_3}]}_{Z_3} =
A^3 B - j^b A^2 B A - j^{b+1} A^2 B A +
j^{2b+1} A B A^2 $$
$$ - j^{b+2} A^2 B A + j^{2b+2} A B A^2
+j^{2b+3} A B A^2 - j^{3b+3} B A^3 = $$
$$ = A^3 B - B A^3 + j^{2b} (j+j^2+j^3) B A^2
- j^b (j+j^2+j^3) A^2 B A = 0 $$
where we have noted b:=grade(B). The result comes from the fact that for any 
grade-1 matrix, the cube $A^3 $ is proportional to the unit matrix and 
therefore commutes with any element $B \in {\cal{A}}$, and because the 
combination $j+j^2+j^3 = j+j^2+1$ is equal to 0. The proof for the grade-2 
derivation is the same.
\newline
\indent
In contrast with what happens in the $Z_2$ -graded Lie algebras, this
derivation does not imply an analogue of the Jacobi identity, because
it {\it is not} a derivation of the $Z_3$ -graded commutator, i.e.
\begin{equation}
{[{[X,Y]}_{Z_3} ,Z]}_{Z_3} + {[{[Y,Z]}_{Z_3} , X]}_{Z_3}
+{[{[Z,X]}_{Z_3} , Y]}_{Z_3} \not = 0 
\end{equation}
An interesting example is provided by derivations of an associative algebra 
which is a ternary generalization of Grassmann algebra. Consider a free 
associative algebra with unit element, on which the following ternary 
relation is imposed:
\begin{equation}
XYZ = j YZX = j^2 ZXY
\end{equation}
Consider the simplest case with one generator only: then the whole algebra 
consists of three elements, $1 , X$ and $X^2 $, because $X^3 = 0 $. If we want 
to define derivations satisfying the $Z_3$-graded Leibniz rule, i.e.
\begin{equation}
\partial (XY) = (\partial X) Y + j X (\partial Y)
\end{equation}
then it is easy to see that only three solutions are possible. They can
be defined explicitly by their action on the three elements of our
algebra:
\begin{equation}
\partial_1 (X) = 1 ; \ \ \partial_1 (X^2) = - j^2 X ; \ \ \partial_1 (1) = 0 ;
\end{equation}
\begin{equation}
\partial_2 (X) = X^2 ; \ \ \partial_2 (X^2) = 0 ; \ \ \partial_2 (1) = 0 ;
\end{equation}
\begin{equation}
\partial_3 (X) = X ; \ \ \partial_3 (X^2) = - j^2 X^2; \ \ \partial_3 (1) = 0
\end{equation}
The derivation $\partial_1 $ is of grade 1 , $\partial_2 $ is of grade 2,
whereas $\partial_3 $  is of grade 0.
The two derivations $\partial_1 $ and $\partial_2 $ do not close under
any binary relations, but they form a simple ternary algebra :
\begin{equation}
\partial_1 \partial_2 \partial_2 + \partial_2 \partial_1 \partial_2 +
\partial_2 \partial_2 \partial_1 = - j^2 \partial_2 ;
\end{equation}
\begin{equation}
\partial_2 \partial_1 \partial_1 + \partial_1 \partial_2 \partial_1 +
\partial_1 \partial_1 \partial_2 = - j^2 \partial_1 .
\end{equation}
We have already seen a similar non-associative ternary algebra realized in 
the set of 3-linear complex forms (Eq.19). Unfortunately, such $Z-3$-graded 
derivations cannot be realized on associative algebras satisfying the above 
permutation ternary rule with more than one independent generator; we can only 
repeat the construction by tensoring some number of identical realizations
introduced above. The $1$-dimensional version of this calculus has been worked
out by W.S.Chung (\cite{Chung})
\newline
\indent
The important fact is that the $Z_3$-graded derivations never close under 
a binary composition rule, but they can produce another derivation under a
ternary composition rule.
\newline
\indent
In the associative algebra of $3 \times 3 $ complex matrices we can also 
consider an exterior differential  $d $ whose {\it cube} vanishes identically,  
$d^3 = 0 $ . Such a differential is defined as a $Z_3$ - graded commutator 
with a matrix from ${\cal{A}}_1$. In next ection we shall use such differentials 
to produce an extended version of usual gauge theories.
\newline
\indent
The associative $Z_3$-graded matrix algebra appears naturally as the algebra
of linear transformations of another nilpotent graded associative algebra,
which is a natural $Z_3$-graded generalisation of Grassmann algebras.
\newline
\indent
By analogy with the $Z_2$-graded Grassmann algebras spanned by the set of
anti-commuting generators, we may introduce an associative algebra spanned
by $N$ generators $\theta^A$, $ A,B = 1,2...N$, whose {\it binary} products
$\theta^A \theta^B$ will be considered as $N^2$  independent quantities, 
whereas we shall impose a {\it ternary} analog of the anti-commutation 
relations:
\begin{equation}
\theta^A \theta^B \theta^C = j  \theta^B \theta^C \theta^A = j^2 \theta^C
\theta^A \theta^B
\end{equation}
\indent
A more precise formulation is to say that the algebra in question is the
universal algebra defined by the above relations.
\newline
\indent
{\tbf Corollary} : The cube of any generator must vanish (because in this 
case the relation (1) amounts to $(\theta^A)^3 = j (\theta^A)^3 = 0$ ; all the 
monomials of order $4$ or higher are identically null (the proof that follows
makes use of the associativity of the postulated product and of the relation
$1$): (the low braces are there just to indicate to which triple of $\theta$'s
the circular permutation is being applied)
\begin{equation}
{\underbrace{\theta^A \theta^B \theta}}^C \theta^D = 
j \theta^B {\underbrace{\theta^C \theta^A \theta}}^D =
j^2 {\underbrace{\theta^B \theta^A \theta}}^D \theta^C = 
\theta^A {\underbrace{\theta^D \theta^B \theta}}^C
= j \theta^A \theta^B \theta^C \theta^D ;
\end{equation}
\indent
therefore, as $1-j \neq 0$, one has $\theta^A \theta^B \theta^C \theta^D = 0$.
\newline
\indent
The dimension of this $Z_3$-graded generalization of  Grassmann algebra
is equal to $ N + N^2 + (N^3 - N)/3$; we may also add a ``neutral'' element
denoted by {\tbf 1} and commuting with all other generators.
\newline
\indent
One can note a dissymmetry between the components of this algebra with the
grades 1 et 2 : as a matter of fact, there are $N$ elements of grade 1 
(the $\theta$'s) and $N^2$ elements of grade 2 ($\theta \theta$).
\newline
\indent
A natural way to re-establish the symmetry is to introduce the set of $N$
``{\it conjugate}'' generators , ${\bar \theta}^A$, of grade 2, that satisfy 
conjugate ternary relations (in which $j$ is replaced by $j^2$):
\begin{equation}
{\bar \theta}^A {\bar \theta}^B {\bar \theta}^C = j^2 {\bar \theta}^B
{\bar \theta}^C {\bar \theta}^A
\end{equation}
\indent
The ternary relation between the $\theta^A$'s can be interpreted as follows:
\newline
$\theta^A {\underbrace{\theta^B \theta}}^C =
j {\underbrace{\theta^B \theta}}^C \theta^A$, which suggests
the following relations between the generators $\theta^A$ and ${\bar \theta}^B$ :
\begin{equation}
\theta^A {\bar \theta}^B  = j {\bar \theta}^B \theta^A , \   \
{\bar \theta}^B \theta^A = j^2 \theta^A {\bar \theta}^B .
\end{equation}
\indent
The $Z_3$-graded algebra so defined can be naturally divided in three parts, 
of grade 0, 1 et 2 respectively, with the dimensions of the sub-spaces of 
grades 1 and 2 being equal: one can write symbolically $A = A_0 + A_1 + A_2$, 
where
\vskip 0.2cm
\hskip 0.2cm
$A_0$ contains: 
{ {\tbf 1}, $\theta^A {\bar \theta}^B , \theta^A \theta^B
\theta^C , {\bar \theta}^A {\bar \theta}^B {\bar \theta}^C , \theta^A \theta^B
{\bar \theta}^C {\bar \theta}^D \ \ {\rm and} \ \ \theta^A \theta^B \theta^C
{\bar \theta}^D {\bar \theta}^E {\bar \theta}^F$ ;
\vskip 0.2cm
\hskip 1cm $A_1$ contains: 
$\theta^A, {\bar \theta}^B {\bar \theta}^C , \theta^A \theta^B
{\bar \theta}^C, \theta^A {\bar \theta}^A {\bar \theta}^B {\bar \theta}^C $,
\vskip 0.2cm
\hskip 0.5cm and $A_2$ contains: ${\bar \theta}^A, \theta^A \theta^B,
\theta^A {\bar \theta}^B {\bar \theta}^C, \theta^A \theta^B \theta^C
{\bar \theta}^D$ .
\vskip 0.2cm
\indent
In the case of usual $Z_2$-graded Grassmann algebras the anti-commutation
between the generators of the algebra and the assumed associativity imply
automatically the fact that {\it all} grade $0$ elements {\it commute} with
the rest of the algebra, while {\it any two} elements of grade $1$ anti-commute.
\newline
\indent
In the case of the $Z_3$-graded generalization such an extension of ternary
and binary relations {\it does not follow automatically}, and must be imposed
explicitly. If we decide to extend these relations to {\it all} elements of
the algebra having a well-defined grade (i.e. the monomials in $\theta$'s
and $\bar{\theta}$'s , then many additional expressions must vanish, e.g.:
\vskip 0.15cm
\hskip 2cm
$\theta^A {\underbrace{\theta^B {\bar \theta}}}^C = 
{\underbrace{\theta^B {\bar \theta}}}^C \theta^A =
\theta^B {\underbrace{{\bar \theta}^C \theta}}^A = 
{\bar \theta}^C \theta^A \theta^B = 0$ ;
\vskip 0.2cm
\noindent
because on the one side, $\theta^A {\bar \theta}^C$ is of grade 0 and commutes 
with all other elements; at the same time, commuting ${\bar \theta}^C$ with
$\theta^A \theta^B$ one gets twice the factor $j^2$, which leads to the
overall factor $ j {\bar \theta}^C \theta^A \theta^B $; this produces a
contradiction which can be solved only by supposing that 
$\theta^A \theta^B {\bar \theta}^C = 0$. The resulting $Z_3$-graded algebra 
contains only the following products of generators:
\begin{equation}
A_1 =  \theta , \lbrace {\bar \theta} {\bar \theta} \rbrace ; \ \ \ \  
A_2 =  {\bar \theta}, \  \ \lbrace \theta \theta \rbrace ; \  \ \  \ 
A_0 = \lbrace \theta {\bar \theta} \rbrace , 
\ \ \lbrace \theta \theta \theta \rbrace , \ \ 
\lbrace {\bar \theta} {\bar \theta} {\bar \theta} \rbrace
\end{equation}
\indent
Let us note that the set of grade $0$ (which obviously forms a sub-algebra
of the $Z_3$-graded Grassmann algebra) contains the products which could
symbolize the only observable combinations of {\it quark fields} in quantum 
chromodynamics based on $SU(3)$-symmetry.
\newline
\indent
If we reorder the basis of our algebra, with all the elements of grade $0$ 
first, next all the elements of grade $1$ and finally the elements of grade 
$2$ in a one-column vector, a general linear transformation that would leave 
these entries in the same order can be symbolized by a matrix whose entries 
have a definite $Z_3$-grade placed as follows:
\begin{equation}
\pmatrix{0 & 2 & 1 \cr 1& 0 & 2 \cr 2 & 1 & 0}  \pmatrix{  0 \cr  1 \cr  2 }
= \pmatrix{0 \cr 1 \cr 2}
\end{equation}
\indent
Under the action of such a matrix, the position of the three grades does not 
change in the resulting column; we shall call such an operator a {\it grade 0} 
matrix. We can introduce two other kinds of matrices that raise all the grades 
by $1$ (resp. by $2$), and call them respectively {\it grade 1} and 
{\it grade 2} matrices:
\begin{equation}
\pmatrix{1 & 0 & 2 \cr 2 & 1 & 0 \cr 0 & 2 & 1} \pmatrix{0 \cr 1 \cr 2} = 
\pmatrix{1 \cr 2 \cr 0}, \ \ {\rm and} \ \ \pmatrix{2 & 1 & 0 \cr 0 & 2 & 1 
\cr 1 & 0 & 2} \pmatrix{0 \cr 1 \cr 2} = \pmatrix{2 \cr 0 \cr 1}
\end{equation}
(the numbers $0, 1, 2$ symbolize the grades of the respective entries in the 
matrices). The notions of {\it hypertrace} and {\it hyperdeterminant} 
generalizing the corresponding notions of {\it supertrace} and 
{\it superdeterminant} been successfully introduced and investigated recently 
by B. Le Roy (\cite{Le Roy}).
If we restrict the character of the matrices, admitting only 
complex-valued matrix elements, then the grades $0, 1$ and $2$ will reduce 
themselves to the following three types of $3 \times 3$-block matrices:
\begin{equation}
\pmatrix{a & 0 &0 \cr 0 & b& 0 \cr 0 & 0 &c} , \  \ \  \ 
\pmatrix{0 & \alpha & 0 \cr 0 & 0 & \beta \cr \gamma & 0 &0 } , 
\  \  \  \  \pmatrix {0 & 0 & \gamma \cr \alpha & 0& 0 \cr 0 & \beta & 0 }
\end{equation}
representing arbitrary matrices with respective grade $0, 1$ and $2$, i.e.
the $Z_3$-graded algebra introduced in the beginning of this Section.
\newline
\indent
Let $\eta$ be a matrix of grade $1$; we can define a formal ``differential'' 
on the $Z_3$-graded algebra of $3 \time 3$ matrices as follows:
\begin{equation}
d B := [ \eta , B ]_{Z_3} = \eta B - j^b B \eta
\end{equation}
\indent
It is easy to show that $d (B C ) = (d B) C + j^b B (d C) $ and that $d^3 = 0$.
The first identity is trivial, whereas the last one follows from the fact that
$\eta^3 = Id$ does commute with all the elements of the algebra.
\newline
\indent
It is also easy to check that $Im (d) \subseteq Ker(d^2) , \ \ {\rm and} \ \ 
Im (d^2) \subseteq Ker (d)$
\vskip 0.4cm
\indent
{\tbf 5. Matrix realization of a $Z_3$-graded gauge theory}
\vskip 0.3cm
\indent
Let ${\cal{A}}$ be an associative algebra with unit element, and let 
${\cal{H}}$ be a free left module over this algebra. Let $A$ be an
${\cal{A}}$-valued 1-form defined on a differential manifold $M$, and let 
$\Phi$ be a function on the manifold $M$ with values in the module ${\cal{H}}$. 
We shall introduce the {\it covariant differential} as usual:
\begin{equation}
D \Phi := d \Phi + A  \Phi  ;
\end{equation}
\indent
If the module is a free one, any of its elements $\Phi$ can be represented
by an appropriate element of the algebra acting on a fixed element of 
${\cal{H}}$, so that one can always write $ \Phi = B \Phi_o $; then the 
action of the group of automorphisms of ${\cal{H}}$ can be translated as the 
action of the same group on the algebra ${\cal{A}}$.
\newline
\indent
Let $U$ be a function defined on $M$ with its values in the group of the 
automorphisms of ${\cal{H}}$. The definition of a covariant differential is 
equivalent with to the requirement $D U^{-1} B = U^{-1} D B $; as in the 
usual case, this leads to the following well-known transformation for the 
connection 1-form $A$ :
\begin{equation}
A \Rightarrow U^{-1} A U + U^{-1} d U  ;
\end{equation}
\indent
But here, unlike in the usual theory, the second covariant differential
$D^2 \Phi$ is not an automorphism: as a matter of fact, we have:
$$D^2 \Phi = d ( d \Phi + A \Phi) + A (d \Phi + A \Phi) = 
d^2 \Phi + d A \Phi + j A d \Phi + A d \Phi + A^2 \Phi $$
the expression containing $d^2 \Phi$ and $d \Phi$ , whereas $D^3 \Phi$ is 
an automorphism, because it contains only $\Phi$ multiplied on the left
by an algebra-valued 3-form:
$$D^3 \Phi = d(D^2 \Phi) + A (D^2 \Phi), {\rm which \ \ gives \ \ explicitly:}$$
$$d( d^2 \Phi + dA \Phi + j A d \Phi + A^2 \Phi ) + A ( d^2 \Phi + 
d A \Phi + j A d \Phi + A d \Phi + A^2 \Phi) $$
With a direct calculus one observes that all the terms containing
$d \Phi$ or $d^2 \Phi$ simplify because of the identity $1+j+j^2=0$, leaving
only
\begin{equation}
D^3 \Phi = ( d^2 A + d ( A^2 ) + A d A + A^3 ) \Phi = (D^2 A) \Phi := \Omega 
\ \ \Phi ;
\end{equation}
Obviously, because $D (U^{-1} \Phi) = U^{-1} (D \Phi) $, one also has: 
\vskip 0.2cm
\centerline{$D^3 (U^{-1} \Phi) = U^{-1} ( D^3 \Phi ) = U^{-1} \Omega \Phi = 
U^{-1} \Omega U U^{-1} \Phi$,} 
\vskip 0.2cm
\noindent
which proves that the 3-form $\Omega$ transforms as usual, 
$\Omega \Rightarrow U^{-1} \Omega U$ when the connection 1-form
transforms according to the law:
$A \Rightarrow U^{-1} A U + U^{-1} dU $.
\newline
\indent
It can be also proved by a direct calculus that the curvature 3-form $\Omega$
does vanish identically for $A = U^{-1} dU $. This computation illustrates
very well the technique of the $Z_3$-graded exterior differential calculus
introduced above: as a matter of fact, one has
\begin{equation}
d( U^{-1} dU ) = d U^{-1} dU + U^{-1} d^2 U ,
\end{equation}
so that the term corresponding to $d^2 A$ gives:
$$d^2 ( U^{-1} dU ) = d^2 U^{-1} d U + j d U^{-1} d^2 U + d U^{-1} d^2 U$$ 
next, the term corresponding to $d(A^2) = d(U^{-1} dU U^{-1} dU ) $ gives 
$$d U^{-1} dU U^{-1} dU + U^{-1} d^2 U U^{-1} dU +
j U^{-1} dU d U^{-1} dU + j U^{-1} dU U^{-1} d^2 U $$
$$ \ \ {\rm whereas} \ \ A dA = U^{-1} dU d U^{-1} dU + U^{-1} dU U^{-1} d^2 U $$
finally, the term $A^3 = U^{-1} dU U^{-1} dU U^{-1} dU$ can be written as 
$- d U^{-1} dU U^{-1} dU$ by virtue of the identity $dU U^{-1} = -UdU^{-1}$ 
which follows from the Leibniz rule applied to 
$U U^{-1} = Id $, i.e. $d(U U^{-1}) = dU U^{-1} + U d U^{-1} = 0$. Using this 
identity whenever possible, and replacing $1 + j$ by $- j^2$, we can reduce 
the whole expression to the following sum of three terms
\vskip 0.2cm
\centerline{$d^2 U^{-1} d U + U^{-1} d^2 U U^{-1} dU - j^2 U^{-1} dU dU^{-1} dU $}
\vskip 0.1cm
whose vanishing does not at all seem obvious. However, it is not very 
difficult to prove that this expression is identically null. First of all, 
it is enough to prove the vanishing of the expression
\vskip 0.2cm
\indent
\hskip 1.5cm
$d^2 U^{-1} + U^{-1} d^2 U U^{-1} - j^2 U^{-1} dU d U^{-1} $ ,
\newline
because all the three terms contain the same factor $d U$ on the right; then,
by multiplying on the left by $U$, we get $U d^2 U^{-1}+d^2 U U^{-1}
-j^2 dU dU^{-1} $
\newline
\indent
At this point let us note that $d^2 (U U^{-1}) = d^2 (Id) = 0$, but then,
according to our $Z_3$-graded Leibniz rule,
$$d^2(U U^{-1}) = d (dU U^{-1} + U dU^{-1}) = d^2 U U^{-1} + j dU dU^{-1} +
dU dU^{-1} + U d^2 U^{-1} $$
\centerline{so that  $ U d^2 U^{-1} + d^2 U U^{-1} = - dU dU^{-1} - 
j dU dU^{-1} $, therefore }
$$-dU dU^{-1} - j dU dU^{-1} - j^2 dU dU^{-1} = (1+j+j^2)\,dU dU^{-1} = 0 $$
\indent
It is amusing to look at one of the simplest possible realizations of this
model in the $Z_3$-graded algebra of $3 \times 3$ complex matrices. In this
case the matrices represent both functions and forms, so that it is easy to
perform the calculus of the $3$-form $\Omega$ and in particular to determine
the condition on the connection $1$-form $A$ leading to $\Omega =0$ , i.e.
defining the set of connections $A$ that are {\it pure gauges}.
\newline
\indent
Let $ A = \pmatrix{0&{\alpha}&0 \cr 0&0&\beta \cr {\gamma}&0&0} $    and
$ \eta = \pmatrix{0&1&0 \cr 0&0&1 \cr 1&0&0} $
Then the condition $\Omega = 0 $ leads to the following equation          
for the coefficients $\alpha, \beta, \gamma $ :
\begin{equation}
(\alpha + \beta + \gamma) + \alpha \beta + \beta \gamma + \gamma \alpha +
\alpha \beta \gamma = 0 ,
\end{equation}
or in a more symmetric form, $(\alpha + 1)(\beta + 1)(\gamma + 1) = 1 $
\newline
\indent
The independent solutions displaying full symmetry under the action
of the permutation group $S_3$ are found by putting
\vskip 0.1cm
\centerline{$ \alpha + 1 = \beta + 1 = \gamma + 1 = 1 \ \ {\rm or} \ \  
j \ \ {\rm or} \ \ j^2 ; $}               
\vskip 0.1cm
Another set of solutions is obtained with
\vskip 0.1cm
\centerline{$ \alpha + 1 = 1 , \ \ \beta + 1 = j , \ \  \gamma + 1 = j^2 $}
\vskip 0.1cm
and by performing all possible permutations of $\alpha, \beta $ and $\gamma$.
The analogue of a gauge transformation acting on the connection form $A$
is given by a usual formula,
\begin{equation}
A \rightarrow A' = U^{-1} A U + U^{-1} d U
\end{equation}
Then it is easy to show that for any $U \in {\cal{A}} $ one has
$\Omega ' = U^{-1} \Omega U $, but the covariant derivative $D$ transforms
properly only if $U \in {cal{A_0}}$; in other case, one has
$ U^{-1} (d + A) U = d + j^u A $.
Let $ U \in {\cal{A}}_0$. Then, 
\begin{equation}
{\rm if} \ \  U = \pmatrix{x&0&0 \cr 0&y&0 \cr 0&0&z} , \ \ {\rm one \ \ has} 
\ \ U^{-1} d U = \pmatrix{0&0&{({z \over x} -1)} \cr {({x \over y})-1}&0&0 \cr
0&{({y \over z}) -1}&0}
\end{equation}
which obviously satisfies the equation (52), therefore corresponding to
the vanishing curvature 3-form $\Omega$. Indeed,
$$ (({z \over x}-1)+1)(({x \over y} -1)+1)(({y \over z}-1)+1) =
{z \over x} {x \over y} {y \over z} = 1 $$
\vskip 0.3cm
\indent
{\tbf 6. $Z_3$-graded Exterior Differential Calculus}
\vskip 0.3cm
The associative algebra of functions we shall be dealing with is generated by $N$ 
generators $\xi^k$, which need not to commute. We shall assume more general
binary relations of the type
\begin{equation}
\xi^i \, \xi^k = \xi^k \, \xi^i \,+\, \epsilon^{ik} = \xi^i \, \xi^k + 
\epsilon^{ki} + \epsilon^{ik},
\end{equation}
so that obviously one must have $\epsilon^{ik} = - \epsilon^{ki}$; the
variables $\xi^k$ span a Heisenberg algebra in an even dimensional case.
\newline
\indent
We postulate that the partial derivatives satisfy the usual Leibniz rule:
\begin{equation}
\partial_i \, \xi^k = \delta^k_i ; \ \ \ \ \partial_i (\xi^k \xi^l) = 
\delta^k_i \, \xi^l + \xi^k \, \delta^l_i , \ \ {\rm and} \ \ \ \ \partial_i 
\,(\epsilon^{km}) = 0,
\end{equation}
and that it holds for a product of any two functions of the variables
$\xi^k$. Our algebra being associative, we have  also $\ \ \partial_i\,
\partial_k = \partial_k \,\partial_i$. Let the definition of the differential 
$d f$ of a function $f$ coincide with the usual one:
\begin{equation}
d f = \frac{\partial \, f}{\partial \xi^k} \, d \xi^k
\end{equation}
\indent
When formally computing higher-order differentials, we shall suppose that
our exterior differential operator $d$ obeys the generalized {\it graded}
Leibniz rule:
\begin{equation}
d\,(\omega \, \phi) = d\,\omega\,\phi + j^{grade(\omega)}\,\omega\,d\,\phi
\end{equation}
and that the grades add modulo $3$ under the associative multiplication of 
exterior forms; the functions are of grade $0$, and the operator $d$ raises 
the grade of any form by $1$, which means that $d \xi^k$ is a 1-form whose 
$Z_3$-grade is $1$ by definition; when applied two times, by iteration, $d$ 
will produce a new entity, which we shall call a {\it 1-form of grade $2$,} 
denoted by $d^2 \xi^k$. Finally, we require that $d^3 = 0$. 
\newline
\indent
We shall suppose that the $Z_3$-graded algebra generated by the forms 
$d \xi^i$ and $d^2 \xi^k$  behaves as a {\it left} module over the algebra of
smooth functions of $\xi$'s. In other words, we shall be able to multiply
the forms $d\xi^i$ , $d^2 \xi^k$, $d\xi^i d\xi^k$ , etc. by the functions 
{\it on the left} only; right multiplication will just not be considered here.
That is why we will write by definition, e.g.                 
\begin{equation}
d (\xi^i \xi^k) := \xi^i d\xi^k + \xi^k d\xi^i
\end{equation}
\indent
Let us note that in contrast to the $Z_2$-graded case, the forms are treated
as a whole, even when multiplied from the left by an arbitrary function;
that means that we cannot identify  e.g. 
$$(\omega_i d\xi^i)(\phi_k d\xi^k) \ \ \ \ {\rm with} \ \ \ \ 
(\omega_i \phi_k) d\xi^i d\xi^k $$ 
\indent
It is equivalent to say that the products of functions by forms are to be 
understood in the sense of tensor products, which is associative, but 
non-commutative.
\newline
\indent
Nevertheless, such an identification can be done for the forms of maximal 
degree (i.e. $3$), which contain the products of the type 
$d \xi^i d \xi^k d \xi^m$ or $d \xi^i d^2 \xi^m$, whose exterior differentials 
vanish irrespective of the order of the multiplication; as a matter of fact,
\begin{equation}
d ((\alpha_i d \xi^i)(\beta_k d \xi^k)(\gamma_m d \xi^m)) = d ((\alpha_i 
\beta_k \gamma_m) d \xi^i d \xi^k d \xi^m)) = 0 .
\end{equation}
\indent
With the $Z_3$-graded Leibniz rule so established, the postulate $d^3 = 0$ 
suggests the ternary and binary commutation rules for the differentials 
$d\xi^i$ and $d^2 \xi^k$. To begin with, consider the differentials of a 
function of the coordinates $\xi^k$, where the "first differential"  $df$ 
coincides with the usual one:
\begin{equation}
df :=  (\partial_i f) d\xi^i \ \ ; \ \ \  \ d^2 f := (\partial_k \partial_i f) 
d\xi^k d\xi^i + (\partial_i f) d^2 \xi^i \ \ ;
\end{equation}
\begin{equation}
d^3 f = (\partial_m \partial_k \partial_i f) d\xi^m d\xi^k d\xi^i + 
(\partial_k \partial_i f) [ d^2 \xi^k d\xi^i + j d\xi^i d^2 \xi^k + 
d\xi^k d^2 \xi^i] ;
\end{equation}
(we recall that the last part of the differential, $(\partial_i f)d^3 \xi^i$, 
vanishes by virtue of the postulate $d^3 \xi^i = 0$).
\newline
\indent
Supposing that partial derivatives commute, exchanging the summation indices 
$i$ et $k$ in the last expression and replacing $1 + j$ by $- j^2$, we arrive 
at the following two conditions that lead to the vanishing of $d^3 f$ :
\begin{equation}
d\xi^m d\xi^k d\xi^i + d\xi^k d\xi^i d\xi^m + d\xi^i d\xi^m d\xi^k = 0 \ \ ; 
\  \ \  \ d^2 \xi^k d \xi^i - j^2  d \xi^i d^2 \xi^k = 0 \ \ .
\end{equation}
\indent
which lead in turn to the following relations:
\begin{equation}
d \xi^i d \xi^k d \xi^m = j d \xi^k d \xi^m d \xi^i , \  \ {\rm and} \  \  
d \xi^i d^2 \xi^k = j d^2 \xi^k d \xi^i .
\end{equation}
\indent
By extending these rules to {\it all} the expressions with a well-defined
grade, and applying the associativity of the $Z_3$-exterior product, we see 
that all the expressions of the type $d \xi^i d \xi^k d \xi^m d \xi^n$ and 
$d \xi^i d \xi^k d^2 \xi^m$ must vanish, and along with them, also the 
monomials of higher order that would contain them as factors. Still, this is 
not sufficient in order to satisfy the rule $d^3 = 0$ on all the forms spanned 
by the generators $d \xi^1$ and $d^2 \xi^k$. It can be proved easily that the 
expressions containing $d^2 \xi^i d^2 \xi^k$ must vanish, too. For example, 
if we take the particular 1-form $\xi^i d \xi^k$ and apply to it the operator 
$d$, we get
\begin{equation}
d (\xi^i d \xi^k) = d \xi^i d \xi^k + \xi^i d^2 \xi^k;  
\end{equation}                                    
\begin{equation}
d^2 (\xi^i d \xi^k) = d^2 \xi^i d \xi^k + (1 + j) d \xi^i d^2 \xi^k = 
d^2 \xi^i d \xi^k - d^2 \xi^k d \xi^i ;
\end{equation}
which leads to $d^3 (\xi^i d \xi^k) = d^2 \xi^i d^2 \xi^k - d^2 \xi^k d^2 \xi^i $; 
then, if we want to keep both the associativity of the "exterior product" and 
the ternary rule for the entities of grade $2$, i.e. 
$d^2 \xi^i d^2 \xi^k d^2 \xi^m = j^2 d^2 \xi^k d^2 \xi^m d^2 \xi^i$, then the 
only solution is to impose $d^2 \xi^i d^2 \xi^k = 0$ and to set forward the 
additional rule declaring that any expression containing {\it four or more} 
operators $d$ must identically vanish.
\newline
\indent
With these rules set, we can check that $d^3 = 0$ on all the forms, whatever
their grade or degree. Let us show how such a calculus works on the example 
of a 1-form $\omega = \omega_k d \xi^k$:
$$d(\omega_k d \xi^k)=(\partial_i \omega_k)d \xi^i d \xi^k+\omega_k d^2 \xi^k 
$$
$$d^2 (\omega_k d \xi^k) = (\partial_m \partial_i \omega_k) d \xi^m d \xi^i 
d \xi^k +(\partial_i \omega_k) (d^2 \xi^i d \xi^k+j d \xi^i d^2 \xi^k)+ 
\partial_i \omega_k d \xi^i d^2 \xi^k $$
exchanging the summation indices $i$ and $k$ in two last terms, using  
$j + 1 = - j^2$ and the commutation relations between $d \xi^k$ and $d^2 \xi^i$, 
we can write
\begin{equation}
d^2 (\omega_k d \xi^k) = (\partial_m \partial_i \omega_k) d \xi^m d \xi^i 
d \xi^k + (\partial_i \omega_k - \partial_k \omega_i) d^2 \xi^i d \xi^k .
\end{equation}
where it is interesting to note how the usual anti-symmetric exterior
differential appears as a part of the whole expression. 
\newline
\indent
The natural symmetry between $j$ et $j^2$ which leads to the possibility of 
choosing one of these two complex numbers as the generator of the group $Z_3$, 
and simultaneous interchanging the r\^oles between the grades $1$ and $2$ 
suggests that we could extend the notion of complex conjugation 
$j \Rightarrow (j)^* =j^2 $, with $((j)^*)^* =j$, to the algebra of $Z_3$-graded 
exterior forms and the operator $d$ itself.
\newline
\indent
It does not seem reasonable to use the ``second differentials'' $d^2 x^i$ as 
the objects conjugate to the ``first differentials'' $d x^i$, because the 
rules of $Z_3$-graded exterior differentiation we have imposed break the 
symmetry between these two kinds of differentials: remember that the products 
$d x^i d x^k$, and $d x^i d x^k d x^m$ are admitted, while we require that 
$d^2 x^i d^2 x^k$ and $d^2 x^i d^2 x^k d^2 x^m $ must vanish.
\newline
\indent
This suggests the introduction of a ``{\it conjugate}'' differential $\delta$
of grade $2$, the image of the differential $d$ under the conjugation $*$, 
satisfying the following conjugate relations:
\begin{equation}
\delta x^i \delta x^k \delta x^m = j^2 \delta x^k \delta x^m \delta x^i , \ \ 
\delta x^i \delta^2 x^k = j^2 \delta^2 x^k \delta x^i \ \ .
\end{equation}
\indent
One notes that $\delta^2 x^k$ is of grade 1 ($2+2 = 4 = 1 (mod \ \ 3)$).
\newline
\indent
All the relations existing between the operator $d$ and the exterior forms 
generated by $d x^i$ and $d^2 x^k$ are faithfully reproduced under the
conjugation $*$ if we consider the $Z_3$-graded algebra generated by the
entities $\delta x^i$ and $\delta^2 x^k$ as a {\it right module} over the
algebra of functions {\it F}({\it M}) , with the operator $\delta$ acting  
{\it on the right} on this module.
\newline
\indent
The rules $d^3 = 0$ and $\delta^3 = 0$ suggest their natural extension:
$d  \delta = \delta  d  = 0 $
\newline
\indent
To conclude, we must stress the following point: although algebraically
consistent, the $Z_3$-graded exterior differential calculus does not yet have
any deep geometrical meaning, unlike the case of usual $Z_2$-grading, with 
the duality between the exterior p-forms and p-cells and complexes,integration, 
and Green-Stokes formulae. In our case, for example, we would like to be able 
to extend these formulae as follows:
\begin{equation}
\int_{C}\, d^2\,\omega = \int_{\partial C}\,d\omega = \int_{\partial^2 C}\,
\omega \, ,
\end{equation}
where the operation $\partial,$ when applied to a ``complex'' $C$, gives a new
complex $\partial C$ of lower dimension, but with the possibility of applying
it twice, and with $\partial^3 C = 0$ for any $C$. 
\vskip 0.4cm
\indent
{\tbf 7. $Z_3$-graded Gauge Theory on Principal and Linear Bundles}
\vskip 0.3cm
\indent
The curvature 3-form $\Omega = d^2 A + d (A^2) + A dA + A^3$ is of grade $0$; 
therefore it must be decomposed along the elements $dx^i dx^k dx^m$ and
$d^2 x^i dx^k$. 
Here is how we can compute its components in a local coordinate system. 
By definition,  $A = A_i dx^i$, so we have:
\vskip 0.1cm
\centerline{$dA = \partial_i A_k dx^i dx^k + A_k d^2 x^k $}
\vskip 0.2cm
\centerline{$d^2 A = \partial_m
\partial_i A_k dx^m dx^i dx^k + \partial_i A_k d^2 x^i dx^k + j \partial_i
A_k dx^i d^2 x^k + \partial_i A_k dx^i d^2 x^k $}
\vskip 0.15cm
\noindent
Replacing $1 + j$ by $ - j^2$, and taking into account $ dx^k d^2 x^i = 
j d^2 x^i dx^k $, we get:
\begin{equation}
d^2 A = ( \partial_m \partial_i A_k ) dx^m dx^i dx^k + (\partial_i A_k -
\partial_k A_i ) d^2 x^i d x^k ;
\end{equation}
$${\rm Then,} \  \ d (A^2) + A d A = d A A + j A dA + A dA = dA A - j^2 A dA , 
{\rm which \ \ leads \ \ to }$$
$$(\partial_i A_k A_m) dx^i dx^k dx^m - j^2 (A_m \partial_i A_k) dx^m dx^i dx^k
+ A_k A_m d^2 x^k dx^m - j^2 A_m A_k dx^m d^2 x^k $$
and due to the relations $dx^m d^2 x^k = j d^2 x^k dx^m$ et $dx^m dx^i dx^k = 
j dx^i dx^k dx^m$,
$$d (A^2) + A dA = (A_m \partial_i A_k - \partial_i A_k A_m) dx^m dx^i dx^k +
(A_k A_m - A_m A_k) d^2 x^k dx^m .$$
\indent
Finally, as $A^3 = A_i A_k A_m dx^i dx^k dx^m $, the curvature 3-form can
be written in local coordinates as follows:
\begin{equation}
\Omega = d^2 A + d (A^2) + A dA + A^3 = \Omega_{ikm} dx^i dx^k dx^m + 
F_{ik} d^2x^i dx^k
\end{equation}
\begin{equation}
{\rm where} \ \ \Omega_{ikm} := \partial_i \partial_k A_m + A_i \partial_k A_m - 
\partial_k A_m A_i + A_i A_k A_m ,
\end{equation}
\begin{equation}
{\rm and} \  \ F_{ik} := \partial_i A_k - \partial_k A_i + A_i A_k - A_k A_i ;
\end{equation}
In $F_{ik}$ we easily recognize the 2-form of curvature of the usual gauge
theories.
\newline
\indent
We know that the expression $F_{ik}$ is covariant with respect to the gauge
transformations; on the other hand, the 3-form $\Omega$ is also covariant;
therefore, the local expression $\Omega_{ijk}$ must be covariant, too.
As a matter of fact, it can be expressed as a combination of covariant
derivatives of the 2-form $F_{ik}$. 
\newline
\indent
In order to find the covariant expression of $\Omega_{ikm}$, it suffices to
recall that due to the particular symmetry of the ternary exterior product
$dx^i dx^k dx^m $,
we can replace $\Omega_{ikm}$ by $\frac{1}{3} (\Omega_{ikm} + j^2  \Omega_{kmi} 
+ j \Omega_{mik})$  and analyze the {\it abelian} case, when this expression
reduces itself to $\Omega_{ikm} = \partial_i \partial_k A_m$. Substituting for 
$\partial_i \partial_k A_m$ 
\newline
the equivalent expression
$\frac{1}{3}(\partial_i \partial_k A_m + j^2 \partial_k \partial_m A_i +
j \partial_m \partial_i A_k )$ and writing it as 
\newline
$\frac{1}{3} ( j (\partial_k \partial_m A_i - \partial_i \partial_k A_m) +
j ( \partial_m \partial_i A_k - \partial_i \partial_k A_m ) ) $ , because
$ 1 = -j - j^2$,
\vskip 0.2cm
we can easily recognize
$\frac{1}{3} ( j \partial_i [ \partial_m A_k - \partial_k A_m] + 
j^2 \partial_k [ \partial_m A_i - \partial_i A_k] )$ ;
\newline
\indent
which in a general non-abelian case must lead to the following expression:
\begin{equation}
\Omega_{ikm} =  \frac{1}{3} [ j D_i F_{mk} + j^2 D_k F_{mi} ] ,
\end{equation}
$${\rm or \ \ equivalently \ \ } \Omega_{ikm} = - \frac{1}{6} [ D_i F_{mk} + 
D_k F_{mi} ] + \frac{ i \sqrt{3}}{6} [D_i F_{mk} - D_k F_{mi} ]$$
\indent
A similar construction in the case of linear connection and curvature is also
easy to perform. Let $\{ {\bf e}_k \}$ denote the set of $N$ independent 
vectors defined at any point of our space (which we suppose locally isomorphic 
to $R^N$), forming a basis. We define the {\it covariant differential} of 
${\bf e}_k$ by means of the {\it covariant derivatives} of the ${\bf e}_k$ 
which define the {\it connection coefficients} $\Gamma^l_{ik}$:
\begin{equation}
\nabla \,{\bf e}_k = \nabla_i \,{\bf e}_k \,d \xi^k \, = \, \Gamma^l_{ik} \,
{\bf e}_l \, d \xi^i
\end{equation}
\indent
Now, when applying this operation second time, we get:
\begin{equation}
\nabla^2 \, {\bf e}_k = \partial_m \, \Gamma^l_{ik} {\bf e}_l \, d \xi^m 
d \xi^i +  \Gamma^l_{ik} ( \nabla_m \, {\bf e}_k ) \, d \xi^m d \xi^i + 
\Gamma^l_{ik} \, {\bf e}_l \,d^2 \, \xi^i 
\end{equation}
which in view of the definition of $\nabla_m {\bf e}_k$ can be written as:
\begin{equation}
\nabla^2 \, {\bf e}_k = \biggl( \partial_m \,\Gamma^l_{ik}
+ \Gamma^l_{mj}\,\Gamma^j_{ik} \, \biggr)\,{\bf e}_l\, d\xi^m d\xi^i + 
\Gamma^l_{ik}\,{\bf e}_l\, d^2 \xi^i
\end{equation}
\indent
In the usual differential geometry we would set by definition $d^2 \xi^k = 0,$
and $d \xi^i \, d \xi^k = - \, d \xi^k \, d\xi^i$, which automatically leads
to the well-known expression
\begin{equation}
\nabla^2 \, {\bf e}_k = R_{mi\,k}^l \, {\bf e}_l \, d\xi^m \wedge d\xi^m
\end{equation}
$$\ \ {\rm with} \ \ \ \  R_{mi\,k}^l = \partial_m \Gamma^l_{ik} - \partial_i 
\Gamma^l_{mk} + \Gamma^l_{mj}\,\Gamma^j_{ik} - \Gamma^l_{ij}\,\Gamma^j_{mk},$$ 
$$\ \ {\rm and} \ \ d \xi^m \wedge d\xi^i = \frac{1}{2} ( d\xi^m \otimes 
d\xi^i - d \xi^i \otimes d\xi^m)$$
\indent
Here we no longer assume $d^2\,\xi^k = 0$ anymore, nor any particular symmetry
of the tensorial product of the differenitals $d\,\xi^k \otimes d\,\xi^m$.
Therefore we must write instead:
\begin{equation}
\nabla^2\,{\bf e}_k = \biggl( \frac{1}{2} \, R_{mi\,k}^l + \frac{1}{2} \,
P_{mi\,k}^l \biggr) \, {\bf e}_l \, d\xi^m \, d\xi^i + \Gamma_{ik}^l \,
{\bf e}_l \, d^2 \, \xi^i
\end{equation}
with a new entity
$$ P_{mi\,k}^l = \partial_m \, \Gamma_{ik}^l + \partial_i\,\Gamma_{mk}^l\,
+ \Gamma_{mj}^l\, \Gamma_{ik}^j + \Gamma_{ij}^l \, \Gamma_{mk}^j $$
\indent
Note that $P_{mi\,k}^l$ does not transform as a tensor under a change of
coordinates, but instead obeys a non-homogeneous transformation law, like the 
connection coefficients. Now, if we calculate the {\it third} covariant 
derivative of ${\bf e}_k$, $\nabla^3\,{\bf e}_k$, we get the following expression:
$$\nabla^3\,{\bf e}_k = \frac{1}{2}\,[\partial_n R_{mi\,k}^l + \partial_n\,
P_{mi\,k}^l \,] \,{\bf e}_l\,d\xi^n\,d\xi^m\,d\xi^i \, + \frac{1}{2}\,
[R_{mi\,k}^l + P_{mi\,k}^l] \,\frac{\partial\,{\bf e}_l}{\partial\,\xi^n}\,
d\xi^n\,d\xi^m\,d\xi^i$$
\begin{equation} +\, \frac{1}{2}\,[\,R_{mi\,k}^l\,+\,P_{mi\,k}^l\,]\,{\bf e}_l\,
[d^2\xi^m d\xi^i +\,p\,d\xi^m d^2\xi^i + d\xi^m d^2\xi^i\,] + \Gamma_{ik}^l\,
{\bf e}_l\,d^3\xi^i
\end{equation}
If we choose $p = j$ and $d^3\xi^k = 0$, this complicated formula simplifies 
to:
$$\nabla^3\,{\bf e}_k = R_{mi\,k}^l\,{\bf e}_l\,d^2\xi^m\,d\xi^i+\frac{1}{2}\,
[\nabla_n\,R_{im\,k}^l\,-\,\nabla_m\,R_{in\,k}^l]\,{\bf e}_l\, d\xi^nd\xi^id\xi^m$$
\begin{equation}
+\,\frac{i\,\sqrt{3}}{2}\,[\nabla_n\,R_{im\,k}^l\,+\,\nabla_m\,R_{in\,k}^l\,]
\,{\bf e}_l\, d\xi^nd\xi^id\xi^m.
\end{equation}
\indent
It is interesting to note that only {\it two} combinations of the covariant 
derivative of $R_{ik\,m}^l$ appear here; as a matter of fact, the third one,
$\nabla_i\,R_{mn\,k}^l$ is linearly dependent by virtue of Bianchi identity.
\newline
\indent
The expression for $\nabla^3\,{\bf e}_k$ in the case of $Z_3$-graded differential
calculus obtained above has also a clear geometrical meaning. In the  usual 
($Z_2$-graded) case, the condition of vanishing of the expression $\nabla^2
{\bf e}_k$ was equivalent with the zero-curvature condition, $R_{im\,k}^l =0$;
here, the vanishing of $\nabla^3 \,{\bf e}_k$ also implies vanishing curvature,
however, another invariant and interesting condition can be formulated, i.e.
$$\nabla^3\,{\bf e}_k \, = \, R_{im\,k}^l\,{\bf e}_l$$
which implies {\it constant} curvature, the condition satisfied in symmetric
spaces.
\vskip 0.4cm
\indent
{\tbf 9. A glimpse at possible future developments}
\vskip 0.3cm
\indent
Classical gauge fields appear as the necessary device that maintains the
covariance of the Dirac equation with respect to {\it unitary} transformations 
of the spinor wave function, $\psi \rightarrow e^{iS}\,\psi$ when $S$ becomes 
a function of the space-time coordinates. Then we replace the free 4-momentum
operator $p_{\mu}$ by its covariant counterpart $p_{\mu} - ie\,A_{\mu}$; 
simultaneously with a gauge transformation $\psi \rightarrow e^{iS}\,\psi$, 
we have (in the simplest, abelian case) $A_{\mu} \rightarrow A_{\mu} + 
\partial_{\mu} S$
\newline
\indent
It is natural to ask what is the generalization of Dirac's equation that would
lead to the modified gauge theory, with curvature $3$-form $\Omega$ introduced
above. The answer is quite obvious: if in the classical case the curvature
$2$-form $F_{\mu \nu}$ containing the {\it first} derivatives of $A$ did appear 
naturally when we diagonalized the Dirac equation, applying once more the
conjugate Dirac operator, here we must introduce a Schr\"odinger-like equation
linear in the momentum operator, only the {\it third power} of which would 
become diagonal.
\newline
\indent
This leads naturally to a ternary generalization of Clifford algebras.
Instead of the usual binary relation defining the usual Clifford algebra,
$$\gamma^{\mu}\,\gamma^{\nu}\,+\,\gamma^{\nu}\,\gamma^{\mu} \,= 2\,g^{\mu \nu} 
\, {\bf 1},\ \ \ \ {\rm with} \ \ g^{\mu \nu}- g^{\nu \mu}\,=\,0 $$
we should introduce its ternary generalization, which is quite obvious
(see also V. Abramov, \cite{Abramov}) :
\begin{equation}
Q^a\,Q^b\,Q^c + Q^b\,Q^c\,Q^a + Q^c Q^b Q^a = 3 \,\eta^{abc}\, {\bf 1},
\end{equation}
where the tensor $\eta^{abc}$ must satisfy 
$ \ \ \, \eta^{abc}\,=\, \eta^{bca}\,= \, \eta^{cab} $
\newline
The lowest-dimensional representation of such an algebra is given by complex
$3 \times 3$ matrices: 
\begin{equation}
Q^1 = \pmatrix{0&1&0\cr 0&0&j \cr j^2&0&0}, \ \ Q^2 = \pmatrix{0&1&0\cr0&0&j^2\cr
j&0&0}, \ \ Q^3 = \pmatrix{0&1&0\cr0&0&1\cr1&0&0}
\end{equation}
These matrices are given the $Z_3$-grade $1$; their hermitian conjugates
$Q^{*a} = (Q^a)^{\dagger}$ are of $Z_3$-grade $2$, whereas the diagonal
matrices are of $Z_3$-grade $0$; it is easy to verify that the so defined
grades add up modulo $3$.
\newline
\indent
The matrices $Q^a$ ($a=1,2,3$) satisfy the ternary relations (49) with 
\newline
$\eta^{abc}$ a totally-symmetric tensor, whose only non-vanishing components 
are 
\newline
$\eta^{111}=\eta^{222}=\eta^{333}= 1 $ , $\eta^{123} = \eta^{231} 
= \eta^{321} =j^2$, and $\eta^{321}=\eta^{213}= \eta^{132}=j$.
\newline
\indent
Therefore, the $Z_3$-graded generalization of Dirac's equation should read:
\begin{equation}
\frac{\partial\,\psi}{\partial\,t} = Q^1 \,\frac{\partial \psi}{\partial \,x} 
+ Q^2 \, \frac{\partial \psi }{\partial y} + Q^3 \, 
\frac{\partial \psi}{\partial z} + B \,m \, \psi
\end{equation}
where $\psi$ stands for a triplet of wave functions, which can be considered
either as a column, or as a grade $1$ matrix with three non-vanishing entries 
$u\, v\, w$, and $B$ is the diagonal $3 \times 3$ matrix with the eigenvalues
$1\,j$ and $j^2$. It is interesting to note that this is possible only with 
{\it three} spatial coordinates.
\newline
In order to diagonalize this equation, we must act three times with the same
operator, which will lead to the same equation of {\it third order}, satisfied
by each of the three components $u,\,v, \,w$, e.g.:
\begin{equation}
\frac{\partial^3\,u}{\partial t^3} = \biggl[ \frac{\partial^3}{\partial x^3}
+ \frac{\partial^3}{\partial\,y^3} + \frac{\partial^3}{\partial z^3} -
\frac{\partial^3}{\partial x \partial y \partial z} \biggr]\,u + m^3\, u
\end{equation}
\indent
This equation can be solved by separation of variables; the time-dependent
and the space-dependent factors have the same structure: 
$$A_1 \,e^{\omega\,t} + A_2 \,e^{j \,\omega\,t} + A_3 e^{j^2 \,\omega\,t},\,
\ \ \ \ B_1\,e^{{\bf k.r}} + B_2\,e^{j\,{\bf k.r}} + B_3\,e^{j^2\,{\bf k.r}}$$ 
and their nine independent  products can be represented in a basis of real 
functions as
\begin{equation}
\pmatrix{ A_{11} \, e^{\omega\,t+{\bf k.r}} & A_{12} \, e^{\omega\,t-\frac{{\bf k.r}}{2}}
\,cos\,\xi& A_{13} \, e^{\omega\,t-\frac{{\bf k.r}}{2}}\,sin\,\xi \cr A_{21}\,
e^{- \frac{\omega\,t}{2}+{\bf k.r }}\,cos\,\tau & A_{22}\,
e^{- \frac{\omega\,t}{2}-\frac{{\bf k.r}}{2}}\,cos\,\tau\,cos\,\xi &
A_{23}\,e^{- \frac{\omega\,t}{2}-\frac{{\bf k.r}}{2}}\,cos\,\tau \sin\,\xi \cr 
A_{31}\,e^{- \frac{\omega\,t}{2}+{\bf k.r}}\,sin\,\tau & A_{32}\,
e^{- \frac{\omega\,t}{2}-\frac{{\bf k.r}}{2}}\,sin\,\tau \,cos\,\xi & A_{33}\,
e^{- \frac{\omega\,t}{2}-\frac{{\bf k.r}}{2}}\,sin\,\tau \,sin\,\xi }
\end{equation}
where $\tau=\frac{\sqrt{3}}{2} \omega\,t$ and $\xi=\frac{\sqrt{3}}{2}{\bf kr}$.
The parameters $\omega, {\bf k}$ and $m$ must satisfy the cubic dispersion
relation:
\begin{equation}
\omega^3 = k_x^3 + k_y^3 + k_z^3 - 3\,k_x k_y k_z + m^3
\end{equation}
\indent
This relation is invariant under the simultaneous change of sign of $\omega$,
${\bf k}$ and $m$, which suggests the introduction of another set of solutions
constructed in the same manner, but with minus sign in front of $\omega$ and
${\bf k}$, which we shall call {\it conjugate} solutions.
\newline
\indent
Although neither of these functions belongs to the space of tempered
distributions, on which a Fourier transform can be performed, their ternary
skew-symmetric products contain only trigonometric functions, depending on
the combinations $2\,(\tau - \xi)$ and $2\,(\tau + \xi)$. As a matter of fact,
not only the {\it determinant}, but also each of the {\it minors} of the
above matrix is a combination of the trigonometric functions only. The same 
is true for the binary products of ``conjugate'' solutions, with the opposite 
signs of $\omega t$ and ${\bf k.r}$ in the exponentials. 
\newline
\indent
This fact suggests that it is possible to obtain via linear combinations of
these products the solutions of {\it second} or {\it first order} differential
esuations, like Klein-Gordon or Dirac equation.
\newline
\indent
Still, the parameters $\omega$ and ${\bf k}$ do not satisfy the proper mass
shell relations; however, it is possible to find new parameters, which are 
linear combinations of these, that will satisfy quadratic relations that may
be intrpreted as a mass shell equation. We can more readily see this if we 
use the following parametrisation: let us put 
$$\zeta = (k_x + k_y + k_z), \ \ \ \ \chi = Re (j k_x + j^2 k_y + k_z),\,\, 
\ \ \eta = Im (j k_x+j^2 k_y+k_z),$$ 
$${\rm and} \ \ \ \ r^2 = \chi^2 + \eta^2 \, \ \ \ \ \phi = Arctg(\eta/\chi).$$
In these coordinates the cubic mass hyperboloid equation becomes
\begin{equation}
\omega^3 - \zeta\,r^2 = m^3
\end{equation}
\indent
Two obvious symmetries can be immediately seen here, the rotation around the
axis $[1,\,1,\,1]$ ($\phi \rightarrow \phi + \delta \phi),$ and simultaneous 
dilatation of $\zeta$ and $r$: 
$$r \rightarrow \lambda \,r, \ \ \zeta \rightarrow \lambda^{-2}\,\zeta$$
\indent
The same relation can be factorized as
$$(\omega + \zeta)\,(\omega^2 - r^2) + (\omega - \zeta)\,(\omega^2 + r^2) =
2\,m^3$$
\indent
We can define a one-dimensional subset of the above $3$-dimensional hypersurface
by requiering
$$ \omega^2 - r^2 = [ \, 2\,m^3 - (\omega - \zeta)\,(\omega^2 + r^2) \,]/
(\omega + \zeta) = M^2 = Const.$$
\indent
If we have {\it three} hypersurfaces (corresponding to the dispersion relations
of three quarks satisfying the $3$-rd order differential equation), which
are embedded in the $12$-dimensional space $M_4 \times M_4 \times M_4$, then
the resulting $3$-dimensional hypersurface defined by the above constrained
applied to each of the three dispersion relations independently will produce
the ordinary mass hyperboloid
$$\omega_1^2 + \omega_2^2 + \omega_3^2 - r_1^2 - r_2^2 - r_3^2 = \Omega^2 -
r_1^2 - r_2^2 - r_3^2 = 3\,M^2 $$
\indent
Another way to achieve a similar result is to observe that we need not 
multiply the solutions of our third-order differential equation {\it pointwise,}
i.e. with the same argument; we should rather multiply the solutions with the
same value of $t$, but with different values of ${\bf k}_a$ et ${\bf r}_b.$ 
Then we must impose supplementary conditions on the parameters $\omega_a ,$
${\bf k}_a$ and ${\bf r}_c$ in order to cancel all real exponentials in these
products. In terms of these variables the resulting constraints amount to
something very close to {\it confinement}, because our solutions will be 
subjected to the conditions of the general type
$${\bf k}_1.{\bf r}_1 - \frac{1}{2} {\bf k}_2 . {\bf r}_2 - \frac{1}{2} 
{\bf k}_3.{\bf r}_3 = 0;$$
which will at the same time factorize the cubic dispersion relation producing
(although not in a unique way) a relativistic mass hyperboloid for certain
linear cominations of $\omega_a$ and ${\bf k}_b$.
\newline
\indent
The solutions of our third-order differential equation do not belong to the
space of tempered distributions and their Fourier transform is not well defined;
also their products can not be represented as inverse Fourier transforms of
the convolution of their Fourier transforms. Nevertheless, as in classical 
field theory, we can do this if their supports are restricted to positive 
frequencies only. Then one can write symbolically the convolution of
three quark field propagators as follows:
$$\frac{1}{\omega^3 - {\bf k}^3 - m^3}\,*\,\frac{1}{\omega^3 - {\bf k}^3 - m^3}
\,*\,\frac{1}{\omega^3 - {\bf k}^3 - m^3}$$
where ${\bf k}^3$ stands for the cubic form $\, \eta^{abc}\,k_a\,k_b\,k_c\,,$ 
and the integral is taken over the cubic hyperboloid (;); to the product of 
wave functions of quark with  anti-quark corresponds {\it one} convolution 
of two factors of this type.
\newline
\indent
According to our hypothesis, the convolution of the Fourier transforms of 
three quark (or anti-quark) propagators should generate the propagator of
the corresponding composed particle, i.e. a fermion, whereas the convolution
of two such propagators should give the propagator of a boson.
\newline
\indent
A simple power counting gives the dimension of the Fourier transform of the
resulting propagator:
\newline
$(-3) \times 3 + 2 \times D = - 9 + 2\,D$ in the first case, and $(-3) \times
2 + D = - 6 + D$ in the second case, where $D$ is the dimension of the 
space- time. It is only when the dimension $D = 4$ that we get the resulting 
propagator of dimension $- 1$  for a ternary combination, and of dimension 
$- 2$ for a binary combination, which is what is observed indeed for 
{\it fermions} and {\it bosons}, i.e. fields obeying the first and second 
order wave equations, respectively.

\vskip.4cm
\indent
{\tbf 10. Postscript}
\vskip 0.3cm
\indent
It was intended that this paper dedicated to Andr\'e Trautman be shorter than
was eventually the case. The reason was that so much of my work, even including
the most recent, can trace its genesis to the early influence of Andr\'e.
Knowing Andr\'e's predilection for chess and chessboards, I hope that the
patchwork of ideas presented above will provide him with an amusement of
finding the resonances and echos of his own ideas.
\vskip 0.4cm
\indent
{\it Acknowledgements.}   
\vskip 0.3cm
{\it Numerous enlightening discussions  with V.Abramov, B.Le Roy, L.Vainerman, 
M.Dubois-Violette, J.Madore, J.Lukierski and P.Kulish are gratefully acknowledged.}


\begin{thebibliography}{100}

\bibitem{Goethe} J.W. Goethe, {\it Faust, erster teil}, (Goethes Werke, Abt.1.
Bd.14.), Weimar, Hermann B\"ohlau ed. (1887); {\it also:} ed. Reclam, Stuttgart
(1992)

\bibitem{Rousseau} J.J. Rousseau, {\it Julie ou la Nouvelle H\'eloise}, 
\'edition pr\'esent\'ee, \'etablie et annot\'ee par H. Coulet, Gallimard, 
Paris (1993).

\bibitem{Fourier} Ch. Fourier, {\it Le socialisme soci\'etaire}, (extrait des
oeuvres compl\`etes, Soci\'et\'e nouvelle de librairie et d'\'edition, Paris
(1903).

\bibitem{Marx} K. Marx, {\it Das Kapital}, (1886)

\bibitem{Lich1} A. Lichnerowicz, Th\'eorie des connexions et Groupes d'holonomie,
Dunod, Paris, (1956)

\bibitem{Trautman1} A. Trautman, Reports in Math.Physics, {\tbf 1},(1971), 29

\bibitem{Trautman2} A. Trautman, {\it Differential Geometry for Physicists},
{\it Bibliopolis}, Napoli, (1984)

\bibitem{Ker1} R. Kerner, Ann.Inst. H.Poincar\'e, {\tbf 9} (2), (1968), 147

\bibitem{Ker2} R. Kerner, Ann.Inst. H.Poincar\'e, {\tbf 34}(4), (1981), 437

\bibitem{Cho} Y.M. Cho and P.G.O. Freund, Phys. Rev. D 12 (1975), 1711

\bibitem{Witten} E. Witten, Nucl. Phys. B 186, (1981), 412

\bibitem{Ker3} R. Kerner, Journ. of Math. Physics, {\tbf 24} (2) (1983), 356

\bibitem{Ker4} R. Kerner, CERN preprint TH 3669 (1983)

\bibitem{Bertrand1} Ch. Bertrand, R. Kerner, Lett.Math.Phys.,{\tbf 18} (1989), 
193

\bibitem{Bertrand2} Ch. Bertrand, R. Kerner, S. Mignemi, Int.Journ.Mod.Phys. A,
{\tbf 7} (31), (1992), 7741

\bibitem{Newton} I. Newton, {\it Principia} edited in: {\it Sir Isaac Newton's
Mathematical Principles of Natural Philosophy and His System of the World},
Andrew Motte, trans. London; reprint: Greenwood, New-York (1969)

\bibitem{Dubois1} M. Dubois-Violette, R. Kerner, J. Madore, Journ.Math.Phys.,
{\tbf 33}, (1990), 312

\bibitem{Dubois2} M. Dubois-Violette, R. Kerner, J. Madore, Journ.Math.Phys.,
{\tbf 33}, (1990), 323

\bibitem{Dubois3} M. Dubois-Violette, J. Madore, R. Kerner, Class. and Quantum
Gravity, {\tbf 8}, (1991), 1077   

\bibitem{Connes1} A. Connes, J. Lott, Nuclear Physics B (Proc.Suppl), {\tbf 18} 
(1990), 29

\bibitem{Coque} R. Coquereaux, G. Esposito-Far\`ese, C. Vaillant, Nucl.Phys. B
{\tbf 353}, (1991), 689 

\bibitem{Madore1} J. Madore, Phys.Lett.A (1992)

\bibitem{Chams} A. H. Chamseddine, G. Felder, J. Fr\"olich, Phys. Lett B,
{\tbf 296}, (1992), 109

\bibitem{Scheck} F. Scheck, Physics Lett. B {\tbf 284} (1992), 303

\bibitem{Ker5} R. Kerner, Comptes Rendus Acad. Sci. Paris, S\'er II, t.312, 
(1991), p.191 - 196.

\bibitem{Ker6} R. Kerner, Journ. Math.Phys. {\tbf 33} (1992), p.403 - 411.

\bibitem{Lawrence} R. Lawrence, {\it ``Algebras and triangle relations''},
in {\it Topological Methods in Field Theory}, ed. by J. Mickelsson  and 
O. Pekoneti, World Scientific, (1992), p. 429 - 447.

\bibitem{Vainerman} L. Vainerman, R. Kerner, Journ. Math. Phys., to appear 
(1996)

\bibitem{Chung} Won-Sang Chung, Journ. Math. Phys. {\tbf 35}(5), (1994), 2497

\bibitem{Ker7} R. Kerner, in {\it ``Generalized Symmetries in Physics''},
p. 375-394, H.D. Doebner, V. Dobrev and A. Ushveridze eds., World Scientific 
(1994).

\bibitem{Le Roy} B. Le Roy, Journ. of Math. Physics, {\tbf 35} (1), (1995),
474

\bibitem{Abramov} V. Abramov, Algebras, Groups and Geometries, {\tbf 12} 
(1995), 201

\end{thebibliography}
\end{document}